\documentclass[twocolumn,pra,showpacs,superscriptaddress]{revtex4}
\pdfoutput=1
\usepackage{setspace} \usepackage{amsfonts,amsmath,amssymb} \usepackage{txfonts}
\usepackage{graphicx, color} \usepackage{hyperref}

\begin{document}

\title{Dynamical thermalization and vortex formation in stirred 2D Bose-Einstein
condensates}

\author{T.  M.  Wright} \affiliation{Jack Dodd Centre for Quantum Technology,
Department of Physics, University of Otago, PO Box 56, Dunedin, New Zealand}

\author{R.  J.  Ballagh} \affiliation{Jack Dodd Centre for Quantum Technology,
Department of Physics, University of Otago, PO Box 56, Dunedin, New Zealand}

\author{A.  S.  Bradley} \affiliation{ARC Centre of Excellence for Quantum-Atom
Optics, School of Physical Sciences, University of Queensland, Brisbane, QLD
4072, Australia} \affiliation{Jack Dodd Centre for Quantum Technology,
Department of Physics, University of Otago, PO Box 56, Dunedin, New Zealand}

\author{P.  B.  Blakie} \affiliation{Jack Dodd Centre for Quantum Technology,
Department of Physics, University of Otago, PO Box 56, Dunedin, New Zealand}

\author{C.  W.  Gardiner} \affiliation{Jack Dodd Centre for Quantum Technology,
Department of Physics, University of Otago, PO Box 56, Dunedin, New Zealand}

\begin{abstract} We present a quantum mechanical treatment of the mechanical
stirring of Bose-Einstein condensates using classical field techniques.  In our
approach the condensate and excited modes are described using a Hamiltonian
classical field method in which the atom number and (rotating frame) energy are
strictly conserved.  We simulate a $T=0$ quasi-2D condensate perturbed by a
rotating anisotropic trapping potential.  Vacuum fluctuations in the initial
state provide an irreducible mechanism for breaking the initial symmetries of
the condensate and seeding the subsequent dynamical instability.  Highly
turbulent motion develops and we quantify the emergence of a rotating thermal
component that provides the dissipation necessary for the nucleation and
motional-damping of vortices in the condensate.  Vortex lattice formation is not
observed, rather the vortices assemble into a spatially disordered \emph{vortex
liquid} state.  We discuss methods we have developed to identify the condensate
in the presence of an irregular distribution of vortices, determine the
thermodynamic parameters of the thermal component, and extract damping rates
from the classical field trajectories.  \end{abstract}


\date{\today}

\maketitle

\section{Introduction} The experimental observation of vortex lattices in
Bose-Einstein condensates (BECs)
\cite{Madison00,Madison01b,Abo-Shaeer01,Abo-Shaeer02,Raman01,Haljan01a,Hodby02}
has stimulated intense theoretical interest and debate over the mechanisms
involved in vortex formation.  While vortex lattices had previously been
observed in superconductors and superfluid Helium, dilute gas BECs offer the
possibility of a quantitative description of the formation process using
tractable theory.  The experiments can be divided broadly into two categories:
(i) A thermal gas rotating in an axially symmetric trap is evaporatively cooled
and condenses into a rotating lattice~\cite{Haljan01a}; (ii) A low temperature
condensate ($T\ll T_\mathrm{C}$ ) is rotationally stirred, and vortices then
nucleate and form a vortex
lattice~\cite{Madison00,Madison01b,Abo-Shaeer01,Abo-Shaeer02,Raman01,Hodby02}.
The first category of experiment is conceptually simplest, and the origin of
dissipative processes is well understood
\cite{Penckwitt02,Williams2002a,Bradley08}.  The second category of experiment
is the subject of this paper, and as observed by a number of groups, has more
complicated phenomenology.  We shall concentrate on a typical, representative
example, in which the condensate is initially in the interacting ground state of
a cylindrically symmetric harmonic trap at a temperature very close to $T=0$.
Stirring is implemented by distorting the trap to elliptical symmetry, and
rotating it at the frequency of the condensate quadrupole mode.  This excites a
dynamical instability in which the condensate is transformed into a turbulent
state, which later evolves into a rotating state containing a regular vortex
lattice.

This process of vortex lattice formation by stirring provides a unique testbed
for dynamical theories of cold bosonic gases, as noted by other authors
\cite{Kasamatsu03}.  Beginning from a well characterised initial state, the
application of a simply defined experimental procedure causes a violent
transformation into a highly excited and disordered state, from which an ordered
state subsequently emerges.  The workhorse of BEC theory, the Gross-Pitaevskii
equation (GPE), can provide a good dynamical description of $T=0$ condensates,
and has given a very good understanding of the initiation of the dynamical
instability \cite{Sinha01,Recati01,Kasamatsu03,Parker06}.  The subsequent
evolution into a highly excited state is clearly beyond the validity regime of
pure GP theory.  Tsubota \emph{et al.} \cite{Tsubota02a} recognised the need for
a new theoretical framework after noting that dissipation is required to evolve
an initially non-rotating GPE ground state into a state in equilibrium with the
stirrer.  They postulated that a rotating thermal cloud develops to provide this
dissipation (see also Ref.  \cite{Penckwitt02}), and modelled its effects by
adding a phenomenological damping term to the GPE~\cite{Choi98a}, later
justified~\cite{Kasamatsu03} on the basis of the formalism of Jackson and
Zaremba \cite{Jackson2001}.  Further calculations employing the GPE were
performed by Parker \emph{et al.} \cite{Parker06b,Parker05a}, and these 2D
simulations gave rise to crystallization of a low energy vortex lattice, in
conflict with earlier simulations of the GPE in 2D \cite{Feder01a,Lundh03} that
did not exhibit lattice formation.  It is of interest to identify the origin of
this discrepancy and to determine the nature of the final state to be expected
in stirring experiments of 2D Bose gases.

In this paper we apply classical field theory, a representation of many-body
physics capable of modelling the dynamical behaviour of both the condensate and
excited states of the matter field.  The formalism for classical field theory
has been developed recently in a number of papers
\cite{Steel98,Davis01,Sinatra01,Sinatra02,Gardiner02,Blakie05,Bradley08}, and it
has been used to describe several phenomena where thermal fluctuations play a
critical role in determining the dynamical behaviour of the system, including:
the condensation transition \cite{Blakie05,Bradley08}, effects of critical
fluctuations on the transition temperature \cite{Davis06}, activation of
vortices in quasi-2D systems \cite{Simula06,Simula2008a,Bisset2008a}, collapse
dynamics of attractive Bose gases \cite{Wuster2007a}, and coherence properties
of spinor condensates \cite{Gawryluk2007a}.  Lobo \emph{et al.} \cite{Lobo04}
have used a version of classical field theory to describe vortex lattice
formation by stirring, and their results indicate that the thermal field
component generated provides the damping necessary for the lattice to form.
Here we will apply an implementation of classical field theory which conserves
particle number and rotating frame energy, and uses a basis choice and numerical
method that has appropriate boundary conditions and is free from numerical
artifacts.  This allows us to provide a detailed and quantitative description of
the thermalization of the condensate, and the nucleation and motional damping of
vortices which is responsible for lattice formation.  We find that the system
relaxes to a state containing a disordered distribution of vortices, in thermal
and rotational equilibrium with a cloud of thermal atoms generated by the
stirring process.

\section{System and equations of motion}\label{sec:SystemEOM} The theory of the
trapped Bose gas can be written in terms of the second quantized Hamiltonian for
the field operator \begin{equation} \hat{\Psi}(\mathbf{x})=\sum_n
\hat{a}_n\phi_n(\mathbf{x}), \end{equation} where the operators satisfy bosonic
commutation relations $[\hat{a}_n,\hat{a}_m^\dag]=\delta_{nm}$, and the modes
$\phi_n(\mathbf{x})$ are any complete orthonormal basis.  In the s-wave
interaction regime the Hamiltonian for the system is given by
\begin{eqnarray}\label{FullH} H&=&\int d^3\mathbf{x}\;
\hat{\Psi}^\dag(\mathbf{x}) H_{\mathrm{sp}} \hat{\Psi}(\mathbf{x})\nonumber\\
 &&+ \frac{U}{2} \int d^3\mathbf{x}\;\hat{\Psi}^\dag(\mathbf{x})
 \hat{\Psi}^\dag(\mathbf{x}) \hat{\Psi}(\mathbf{x})\hat{\Psi}(\mathbf{x})
\end{eqnarray} where $U=4\pi\hbar^2 a/m$, $a$ is the scattering length and $m$
is the atomic mass.  The single particle Hamiltonian is generally of the form
$H_{\mathrm{sp}}=H^0_{\mathrm{sp}}+V_\epsilon(\mathbf{x})+\delta
V(\mathbf{x},t)$ where \begin{equation}\label{Hsp}
H^0_{\mathrm{sp}}=\frac{-\hbar^2\nabla^2}{2m}+V_0(\mathbf{x})-\Omega L_z,
\end{equation} and we have expressed the single particle Hamiltonian in a frame
rotating at angular frequency $\Omega$ about the $z$-axis and decomposed the
external trapping potential into \begin{equation}\label{Vdef}
V(\mathbf{x},t)=V_0(\mathbf{x})+V_\epsilon(\mathbf{x})+\delta V(\mathbf{x},t).
\end{equation} We have singled out $V_0(\mathbf{x})$ in Eqs.~(\ref{Hsp}),
(\ref{Vdef}) as it will be used below to define the single particle basis for
the system.  $V_\epsilon(\mathbf{x})$ is the remaining time independent
potential, and any time dependent potentials present are represented by $\delta
V(\mathbf{x},t)$.

\subsection{The system}\label{sec:System} We consider a condensate initially at
temperature $T=0$ and confined in a cylindrically symmetric harmonic trap with
trapping potential \begin{equation}\label{V0def} V_0(\mathbf{{\normalcolor
x}})=\left(m/2\right)\left\{
\omega_r^{2}\left(x^{2}+y^{2}\right)+\omega_{z}^{2}z^{2}\right\}.
\end{equation} The trap is deformed at time $t=0$ into an ellipse in the
$xy$-plane which rotates around the $z$-axis with angular frequency $\Omega$.
In the frame rotating at frequency $\Omega$, $\delta V=0$ and the trapping
potential takes the form \begin{equation} V(\mathbf{
x},t)=V_0(\mathbf{x})+V_\epsilon(\mathbf{x}), \label{eq:Vtime}\end{equation}
 where \begin{equation}
V_\epsilon(\mathbf{x})=\epsilon m\omega_r^2(y^2-x^2).  \end{equation} In this
paper we set the trap eccentricity to $\epsilon\equiv 0.025$ which is sufficient
to excite the quadrupole instability at $\Omega\sim
\omega_r/\sqrt{2}$~\cite{Sinha01,Recati01}.  In this work we will always drive
the system within the hydrodynamic resonance, choosing $\Omega\equiv
0.75\omega_r$.  The sudden turn-on of the rotating potential at fixed $\epsilon,
\Omega$, means that the rotating frame energy (the quantum average of
Eq.~(\ref{FullH})) is a conserved quantity of the motion.

For computational reasons we now restrict our attention to an effective two
dimensional system.  Such systems have been experimentally realised (e.g.,
\cite{Stock05}) by setting the trapping frequency $\omega_{z}$ sufficiently high
that no modes are excited in the $z$ direction.  A 2D description of our system
is thus valid provided that \cite{Petrov00} \begin{equation}
	 \mu + kT \ll \hbar\omega_{z},
\end{equation}
 where $\mu$ and $T$ are the system chemical potential and temperature
 respectively.  Provided that the oscillator length $l_z = \sqrt{\hbar/
 m\omega_z}$ corresponding to the $z$ confinement greatly exceeds the s-wave
 scattering length $a$, the scattering is well described by the usual 3D contact
 potential \cite{Petrov00}, and we obtain an effective two-dimensional equation
 in which the collisional interaction parameter is simply rescaled to
 $U_\mathrm{2D} = 2\sqrt{2\pi} \hbar^2 a/ml_z$ \cite{Bradley08, Petrov00}.  In
 this work we consider $^\mathrm{23}$Na atoms confined in a strongly oblate
 trap, with trapping frequencies $(\omega_r,\omega_z)=2\pi\times(10,2000)
 \mathrm{rad/s}$.  The s-wave scattering length is $a=2.75\mathrm{nm}$ and the
 oscillator length $l_z = 468\mathrm{nm}$.  Having chosen appropriate
 parameters, in the next section we make the dimensional reduction formally
 rigorous by introducing an energy cutoff into the theory which excludes all
 excited $z$-axis modes.

\subsection{Classical field theory\label{sec:Formalism}} In projected classical
field theory a cutoff energy $E_\mathrm{R}$ is introduced to separate the system
into a low energy region ({\bf L}) with single particle energies satisfying
$\epsilon_n<E_\mathrm{R}$, and a high energy region ({\bf H}) containing the
remaining modes.  Following \cite{Gardiner02,Gardiner03,Bradley08} we shall
refer to these regions as the condensate and non-condensate bands, respectively.
This separation is motivated by the expectation that low energy modes will be
significantly occupied.  Where possible it is most convenient to carry out the
separation in the basis which diagonalizes the single particle Hamiltonian of
the system since at high energies ($\sim E_\mathrm{R}$) the many-body
Hamiltonian (Eq.~\ref{FullH}) is approximately diagonal in this basis (i.e., at
such an energy a cutoff imposed at a single particle energy is approximately
parallel to one imposed at an energy of the interacting system).

For certain potentials an appropriate numerical quadrature method exists which
also greatly expedites simulations.  Any numerical representation of field
theory on a computer is also subject to a restriction of the modes.  In what
follows we provide an explicit connection between the cutoff in our theory and
our numerical implementation.

We begin by introducing the projected field operator \begin{eqnarray}
\hat{\psi}(\mathbf{x})&\equiv& \sum_{n \in
\mathbf{L}}\hat{a}_n\phi_n(\mathbf{x})\nonumber\\ &=&{\cal
P}\hat{\Psi}(\mathbf{x}), \end{eqnarray} where the projector is defined as
\begin{equation}\label{Pdef} {\cal P}f(\mathbf{x})\equiv\sum_{n \in
\mathbf{L}}\phi_n(\mathbf{x})\int d^3\mathbf{y}\;
\phi_n^*(\mathbf{y})f(\mathbf{y}).  \end{equation} The summation is over all
modes satisfying $\epsilon_n\leq E_\mathrm{R}$, where
$H^{0}_{\mathrm{sp}}\phi_n(\mathbf{x})=\epsilon_n\phi_n(\mathbf{x})$ defines the
modes in terms of a time independent single particle Hamiltonian in the rotating
frame.  Choosing $E_\mathrm{R} < \hbar\omega_z$, all excited $z$-axis modes are
excluded from this set.  In the s-wave limit, the effective low energy
Hamiltonian for the system is then given by \begin{equation}\label{Heff}
H_{\mathrm{eff}}=\int d^2\mathbf{x}\; \hat{\psi}^\dag(\mathbf{x})
H_{\mathrm{sp}} \hat{\psi}(\mathbf{x}) +\frac{U_\mathrm{2D}}{2}
\hat{\psi}^\dag(\mathbf{x}) \hat{\psi}^\dag(\mathbf{x})
\hat{\psi}(\mathbf{x})\hat{\psi}(\mathbf{x}) \end{equation} We now obtain a
classical theory by making the replacement $\hat{\psi}(\mathbf{x})\to
\psi(\mathbf{x})$ in $H_{\mathrm{eff}}$ to give \begin{equation}\label{HCF}
H_{\mathrm{CF}}=\int d^2\mathbf{x}\; \psi^*(\mathbf{x}) H_{\mathrm{sp}}
\psi(\mathbf{x}) +\frac{U_\mathrm{2D}}{2} |\psi(\mathbf{x})|^4.  \end{equation}
This approach is based on the high occupation condition, where it is known that
classical behaviour is recovered for highly occupied bosonic modes since
commutators become relatively unimportant~\cite{Davis01}.  We then obtain the
equation of motion for the classical field via projected functional
differentiation of the classical field Hamiltonian~\cite{Gardiner03}:
\begin{equation} i\hbar\frac{\partial\psi(\mathbf{x})}{\partial
t}=\frac{\bar{\delta}H_{\mathrm{CF}}}{\bar{\delta}\psi^*(\mathbf{x})}.
\end{equation} The resulting equation of motion, the \emph{projected}
Gross-Pitaevskii equation (PGPE) \begin{equation} \label{pgpe}
i\hbar\frac{\partial\psi(\mathbf{x})}{\partial t} = {\cal
P}\left\{\left(H_{\mathrm{sp}}+U_\mathrm{2D}|
\psi(\mathbf{x})|^2\right)\psi(\mathbf{x})\right\}, \end{equation} satisfies
several identities governing the evolution of total number $N=\int
d^2\mathbf{x}\;|\psi(\mathbf{x})|^2$, rotating frame energy $H_{\mathrm{CF}}$,
and angular momentum $\overline{L_z}$~\cite{Bradley05}:
\begin{widetext}\vspace{-.5cm} \begin{eqnarray} \label{Ndot}\frac{d
N}{dt}&=&0,\\ \label{Lzdot}\frac{d \overline{L_z}}{dt} &=&
-\frac{i}{\hbar}\overline{ L_z V(\mathbf{x})} + \frac{2}{\hbar}{\rm
Im}\left[\int d^2\mathbf{x} \;\psi^*(\mathbf{x})
\left(V_\epsilon(\mathbf{x})+U_\mathrm{2D}|\psi(\mathbf{x})|^2\right){\cal Q}
\left\{L_z \psi(\mathbf{x})\right\}\right],\\ \label{Edot}\frac{d
H_{\mathrm{CF}}}{dt} &=& -\frac{2\Omega}{\hbar}{\rm Im} \left[\int
d^2\mathbf{x}\;\psi^*(\mathbf{x})
\left(V_\epsilon(\mathbf{x})+U_\mathrm{2D}|\psi(\mathbf{x})|^2\right){\cal Q}
\left\{L_z \psi(\mathbf{x})\right\}\right],\end{eqnarray} \end{widetext} where
${\cal Q}=1-{\cal P}$ is the projector orthogonal to ${\cal P}$, and the bar
represents spatial averaging: $\overline{A}=\int
d^2\mathbf{x}\;\psi^*(\mathbf{x})A\psi(\mathbf{x})$.  The terms involving the
imaginary part of an integral are boundary terms arising in the projected
classical field theory.  Although we work in the frame where $\delta
V(\mathbf{x},t)\equiv 0$, Eqs.  (\ref{pgpe})-(\ref{Edot}) hold for any
partitioning of the time independent potential into additive parts
$V_0(\mathbf{x})$ and $V_\epsilon(\mathbf{x})$ and therefore the precise choice
of $V_0(\mathbf{x})$ defining our projector in Eq.  (\ref{Pdef}) is in principle
arbitrary.  However, in order that we recover the formal properties of the
continuous classical field theory, the boundary terms in Eqs.~(\ref{Lzdot}),
(\ref{Edot}) must be made to vanish.  This is ensured by expanding
$\psi(\mathbf{x})$ in eigenstates of $L_z$, which is equivalent to requiring the
projection operator to have rotational symmetry: $[{\cal P},L_z]\equiv [{\cal
Q},L_z]=0$.  We then have ${\cal Q}\left\{L_z \psi(\mathbf{x})\right\}\equiv 0$
and we see that our choice of $V_0(\mathbf{x})=m \omega_r^2r^2/2$ given in
Sec.~\ref{sec:System} generates the appropriate rotating frame classical field
theory.  The remaining potential $V_\epsilon(\mathbf{x})=\epsilon
m\omega_r^2(y^2-x^2)/2$ is not diagonal in our representation, but since it is
only a weak perturbation the many-body Hamiltonian will remain approximately
diagonal in the eigenstates of $H^0_{\mathrm{sp}}$ near the cutoff energy
$E_\mathrm{R}$.  With this choice of basis, we recover $dN/dt\equiv d
H_{\mathrm{CF}}/dt=0$, and $d\overline{L_z}/dt=(-i/\hbar)\overline{L_z
V_\epsilon(\mathbf{x})}$ as required.

We emphasize that for certain potentials there exists an exact numerical
quadrature method for evolving the PGPE which implements the projection operator
to machine precision for a given energy cutoff.  Such a method exists for our
choice of $V_0(\mathbf{x})$~\cite{Bradley08} and consequently there is an exact
numerical correspondence between the formal properties of the classical field
theory expressed in Eqs.  (\ref{Ndot})-(\ref{Edot}) and the results of PGPE
simulation.  PGPE evolution via numerical quadrature in the basis of $L_z$
eigenstates is thus free of boundary term artifacts, and we have obtained an
appropriate Hamiltonian classical field theory in the rotating frame.

\subsection{Projected truncated Wigner method}\label{sec:TWA} We have thus far
confined our discussion to pure classical field theory without recourse to
phase-space methods.  Having established the appropriately conserving formalism
we also include vacuum noise in our initial conditions, following the
prescription of the truncated Wigner (TW) method~\cite{Steel98}.  The classical
field $\psi$ is formally related to the field operator $\hat{\psi}$ by the
Wigner function phase-space representation
\cite{Steel98,Gardiner02,Gardiner03,Bradley06a,Sinatra01}, and thus even at zero
temperature the initial state of $\psi$ in any trajectory contains a
representation of the vacuum fluctuations in the form of classical (complex)
Gaussian noise of mean population equal to $\frac{1}{2}$-quantum per mode.  It
is of concern, then, to choose appropriate basis modes to which the noise is
added (not to be confused with the basis modes in which the system is
propagated), so as to faithfully represent the ground state of the many-body
system \cite{Steel98}.

While higher order approaches which take account of the mutual interaction
between the condensate and Bogoliubov modes exist~\cite{note1}, in this work our
primary aim is to include the possibility of spontaneous processes in the
initial condition.  It therefore suffices to populate the Bogoliubov modes
orthogonal to the condensate mode with noise~\cite{Gardiner97,Castin98,Steel98}.
The initial state is constructed from the GP ground state of the symmetric trap
in the lab frame, $\phi_0(\mathbf{x})$, as \begin{equation}
\psi(\mathbf{x})=\phi_0(\mathbf{x})+\sum_j
u_j(\mathbf{x})\alpha_j+v_j(\mathbf{x})\alpha_j^*, \end{equation} where
$(u_j,v_j)$ are the Bogoliubov modes orthogonal to $\phi_0(\mathbf{x})$,
$\alpha_j=\sqrt{n_j+1/2}(\xi_j+i\eta_j)/\sqrt{2}$, and for our system the
thermal population $n_j$ is zero.  The independent Gaussian distributed
variables satisfy $\overline{\xi_j}=\overline{\eta_j}=\overline{\eta_i\xi_j}=0$
and $\overline{\xi_i\xi_j}=\overline{\eta_i\eta_j}=\delta_{ij}$.

Population of the quasiparticle basis constructed in this manner ensures that
all surface modes of the condensate (that are resolvable within the condensate
band) are effectively seeded by noise.  The dynamically unstable excitations of
the condensate in the presence of the rotating trap anisotropy are among those
seeded, and the noise introduced here thus plays a role entirely analogous to
that played by vacuum electromagnetic field fluctuations in triggering the
spontaneous decay of a two-level atom \cite{Scully97}.  All symmetries of the
mean-field state are thus broken \cite{Sinatra00} \emph{before} the exponential
growth of non-condensed field density \cite{Castin97} occurring during the
dynamical instability.  By contrast an exact GP ground-state of the isotropic
trap possesses circular rotational symmetry, and thus the \emph{exact} evolution
of the state under the GPE with an elliptic potential would retain unbroken
twofold rotational symmetry for all time.

In any practical calculation, with finite-precision arithmetic, numerical error
eventually leads to the breaking of this symmetry, and researchers propagating
the GPE from such an initial state have supplemented their numerics with
additional procedures to speed this process \cite{Penckwitt02, Parker05a,
Parker06b}.  In this work we consider an irreducible source of symmetry
breaking: vacuum fluctuations.

There are also more subtle reasons for using a projective method and for our
choice of basis: (i) \emph{Preservation of symmetries}.--- It is natural in
modelling the field theory of particles interacting in free space, to consider
the problem on a discrete spatial lattice~\cite{Smit02}, effectively imposing a
\emph{momentum} cutoff on the system.  However, the inclusion of an external
confining potential breaks the translational symmetry of the system, and
momentum is no longer a well-conserved quantity appropriate for defining a
cutoff.  A projective method allows us to implement a finite dimensional field
theory which best respects the remaining symmetries of the confined system
(e.g., the conservation of total field energy).  (ii) \emph{Phase space
discretization}.--- In modelling a homogeneous system on a discrete lattice, one
expects to regain the full, continuum field theory as the lattice spacing tends
to zero (i.e., as the momentum cutoff is raised to infinity).  However, in a
confined system, a \emph{natural} discretization of the field theory is imposed
by the quantization of energy levels in the confining potential.  A cutoff
defined in energy imposes both short-wavelength (ultraviolet) and
long-wavelength (infrared) cutoffs \emph{consistently}, and moreover, ensures
that the number of modes spanning the intervening phase-space is optimally close
to the true number present in the interacting system.  Furthermore, the choice
of an energy cutoff allows us to choose a phase-space appropriate to the
\emph{rotating frame}--the frame in which final equilibrium is achieved.  This
has immediate consequences for the distribution of thermal energy in the system.
A non-optimal discretization, such as a Cartesian grid or planewave model of a
trapped system, leads to somewhat arbitrary results for the number of modes, and
consequently the temperature of the final equilibrium state.  As we will see,
our equilibrium states still have a cutoff dependent temperature, but the
numerical prefactor arising from our choice of basis is close to optimal.  A
complete resolution of this problem requires a more general theory that includes
above-cutoff corrections \cite{note2}.  (iii) \emph{Boundary conditions}.---
Methods commonly used such as Fourier spectral methods~\cite{Lobo04}, and the
Crank-Nicholson method~\cite{Parker_PhD} can suffer from pathologies arising
from periodic boundary conditions, wave vector aliasing~\cite{Press92}, or
discretization error in the calculation of derivatives; both methods are
non-projective and as such do not impose a consistent energy cutoff.  For
example, a model of the homogeneous Bose gas implemented using the Fourier
spectral method will always suffer aliasing issues for any modes with momentum
greater than half the maximum representable on the grid~\cite{note3}.  These
issues are resolved by the use of a projective method.  The system boundary is
defined by the formally energy-restricted theory, and the numerically propagated
basis modes and their coupling are unambiguously defined in direct
correspondence to this theory.

\section{Numerical implementation}\label{sec:Numerical} We introduce now the
dimensionless units we use in our implementation of the PGPE.  We use harmonic
oscillator units ($\{\bar{r},\bar{\omega},\bar{E},\bar{t}\}$) related to SI
units ($\{r,\omega,E,t\}$) by the expressions $r = \bar{r}r_0 $, $\omega =
\bar{\omega}\omega_r$, $E = \bar{E}\hbar\omega_r$, $t = \bar{t}\omega_r^{-1}$,
where the radial oscillator length $r_0 = \sqrt{\hbar/m\omega_r}$.  In reporting
our results we shall often refer to times in units of the trap cycle
(abbreviated $\mathrm{cyc.}$), i.e.  the period of the radial oscillator
potential, $T_\mathrm{osc}=2\pi/\omega_r$.  The energy cutoff $E_\mathrm{R}$
separating the condensate and non-condensate bands is conveniently expressed in
terms of $\mu_\mathrm{i}$, and is typically chosen to be $E_\mathrm{R} \sim
2-3\hbar\mu_\mathrm{i}$.  This choice is made to be high enough so that: (i)
there are sufficient available states to give a reasonable representation of
thermalized atoms, and; (ii) the excited states at the cutoff are only weakly
affected by the mean-field interaction.  However, $E_\mathrm{R}$ must not be too
large: the energy scale of our low energy effective field theory must be well
below that associated with the effective range of the interatomic potential, so
that the contact s-wave potential remains valid \cite{Gardiner03}.  Furthermore,
the number of modes in our theory must be kept as low as possible, so that the
added vacuum noise does not dominate the real population \cite{Sinatra02}.  We
examine the sensitivity of our system to the choice of cutoff in
Sec.~\ref{app:AppendixB}.

\subsection{PGPE method} Here we briefly discuss our implementation of the PGPE
in terms of rotating frame harmonic oscillator states.  A more thorough
discussion of the method is presented in \cite{Bradley08}.  In our dimensionless
units the PGPE (Eq.~(\ref{pgpe})) becomes \begin{equation}\label{pgpeBar}
	i\frac{\partial\bar{\psi}}{\partial \bar{t}} =
	\mathcal{P}\{[-\frac{\bar{\nabla}^2}{2} +
	\frac{\bar{r}^2}{2}(1-\epsilon\cos2\theta) +
	i\bar{\Omega}\partial_\theta + \lambda|\bar{\psi}|^2]\bar{\psi}\},
\end{equation} where the dimensionless effective interaction strength $\lambda =
U_\mathrm{2D}/r_0^2\hbar\omega_r$.  We proceed as in \cite{Bradley08} by
expanding $\bar{\psi}$ as \begin{equation}\label{eq:expansion}
	\bar{\psi}(\bar{\mathbf{x}},\bar{t}) = \sum_{\{n,l\}}
	c_{nl}(\bar{t})\bar{Y}_{nl}(\bar{r},\theta),
\end{equation} over the Laguerre-Gaussian modes
\begin{equation}\label{eq:Laguerre-Gaussian}
	\bar{Y}_{nl}(\bar{r},\theta) = \sqrt{\frac{n!}{\pi(n+|l|)!}}
	e^{il\theta}\bar{r}^{|l|}e^{-\bar{r}^2/2}L_n^{|l|}(\bar{r}^2),
\end{equation} which diagonalize the isotropic single particle (non-interacting)
Hamiltonian $\bar{H}_\mathrm{sp}^0$ \begin{equation}
	\bar{H}_\mathrm{sp}^0\bar{Y}_{nl}(\bar{r},\theta) =
	[-\frac{\bar{\nabla}^2}{2} + \frac{\bar{r}^2}{2} +
	i\bar{\Omega}\partial_\theta]\bar{Y}_{nl}(\bar{r},\theta) =
	\bar{E}_{nl}^\Omega \bar{Y}_{nl}(\bar{r},\theta),
\end{equation} with eigenvalues $\bar{E}_{nl}^\Omega = 2n + |l| - \bar{\Omega} l
+ 1$.  The energy cutoff defined by the projector $\mathcal{P}$ requires the
expansion of Eq.~(\ref{eq:expansion}) to exclude all terms except those in
oscillator modes $\bar{Y}_{nl}(\mathrm{r},\theta)$ for which
$\bar{E}_{nl}^\Omega \leq \bar{E}_\mathrm{R}$, i.e.  those modes satisfying
\begin{equation}\label{eq:band_spectrum}
	2n+|l| - \bar{\Omega} l + 1 \leq \bar{E}_\mathrm{R}.
\end{equation} This yields the equation of motion
\begin{equation}\label{eq:PGPE_coeff_eom}
	i\frac{\partial c_{nl}}{\partial \bar{t}} = \bar{E}_{nl}^\Omega c_{nl} +
	\lambda F_{nl}(\bar{\psi}) + H_{nl}(\bar{\psi}).
\end{equation} The projection of the GPE nonlinearity
\begin{equation}\label{eq:nonlinearity_quadrature}
	F_{nl}(\bar{\psi}) = \int_0^{2\pi} d\theta \int_0^\infty \bar{r}d\bar{r}
	\bar{Y}^*_{nl}(\bar{r},\theta)|
	\bar{\psi}(\bar{r},\theta)|^2\bar{\psi}(\bar{r},\theta),
\end{equation} is evaluated as in \cite{Bradley08}, and this equation of motion
(Eq.~(\ref{eq:PGPE_coeff_eom})) differs only from the deterministic portion of
the SGPE presented there by the inclusion of the projection of the trap
anisotropy \begin{equation}\label{eq:anisotropy}
	H_{nl}(\bar{\psi}) = -\frac{\epsilon}{2}\int_0^{2\pi} d\theta
	\int_0^\infty \bar{r} d\bar{r} \bar{Y}_{nl}^*(\bar{r},\theta)
	\bar{r}^2\cos(2\theta) \bar{\psi}(\bar{r},\theta).
\end{equation} We implement the calculation of this term by means of a combined
Gauss-Laguerre-Fourier quadrature rule similar to that introduced in
\cite{Bradley08} to calculate the $F_{nl}(\bar{\psi})$, and refer the reader to
Appendix \ref{app:AppendixA} for the details.

The PGPE as written in Eq.~(\ref{eq:PGPE_coeff_eom}) immediately lends itself to
numerical integration using the interaction picture method \cite{Gardiner04,
Ballagh00b} with respect to the isotropic single particle Hamiltonian
$H_\mathrm{sp}^0$.  Furthermore the absence of noise terms allows us to use the
adaptive-RK variant \cite{Davis_DPhil} of the method.  This affords us explicit
control over the level of truncation error arising during the integration.  In
practice we choose the accuracy of our integrator to be such that the relative
change in field normalisation is $\lesssim 4\times10^{-9}$ per step taken by the
integrator, and the simulations presented here all have total fractional
normalisation changes $\lesssim 2\times10^{-4}$ over their duration ($\sim 10^4$
trap cycles).  Most importantly, the change in rotating frame \emph{energy} is
of magnitude commensurate with that of that change in normalisation, so that the
truncation error represents a loss of population from the system as a whole,
rather than a preferential removal of high-energy components as in
\cite{Parker06b}, and the relaxation of the condensate band field is thus due to
internal damping processes only.

\section{Simulation Results}\label{sec:Results} We present here the results of a
simulation with initial chemical potential $\mu_\mathrm{i} = 14\hbar\omega_r$,
whose response to the stirring is representative of systems throughout the range
$4\hbar\omega_r \leq \mu_\mathrm{i} \leq 20 \hbar\omega_r$.  Using the criteria
discussed in Sec.~\ref{sec:Numerical} we set the condensate band cutoff
$E_\mathrm{R} = 3\mu_\mathrm{i} =42\hbar\omega_r$.  The upper limit of the range
investigated ($ \mu_\mathrm{i} \sim 20 \hbar\omega_r $) is set by computational
limitations: at the fixed rotation frequency $\Omega=0.75\omega_r$ the size of
the Gauss-Laguerre (GL) basis scales as $\mathcal{M} \sim E_\mathrm{R}^2$, and
the corresponding computational load scales as $\mathcal{O}(E_\mathrm{R}^3)$
\cite{Bradley08}, so that simulations rapidly become numerically expensive.  For
the cutoff $E_\mathrm{R}=42\hbar\omega_r$ the GL basis consists of $\mathcal{M}
= 2028$ modes.  At the lower end of the range ($\mu_\mathrm{i}\lesssim
10\hbar\omega_r$) the simulation results differ somewhat both in the
thermalization process leading to vortex nucleation, and the behaviour of the
vortex array, due to the reduced mean-field effects and small numbers of
vortices respectively.  These differences will be discussed in
Sec.~\ref{subsec:vary_mu}.

\begin{figure*}
	\includegraphics[width=0.9\textwidth]{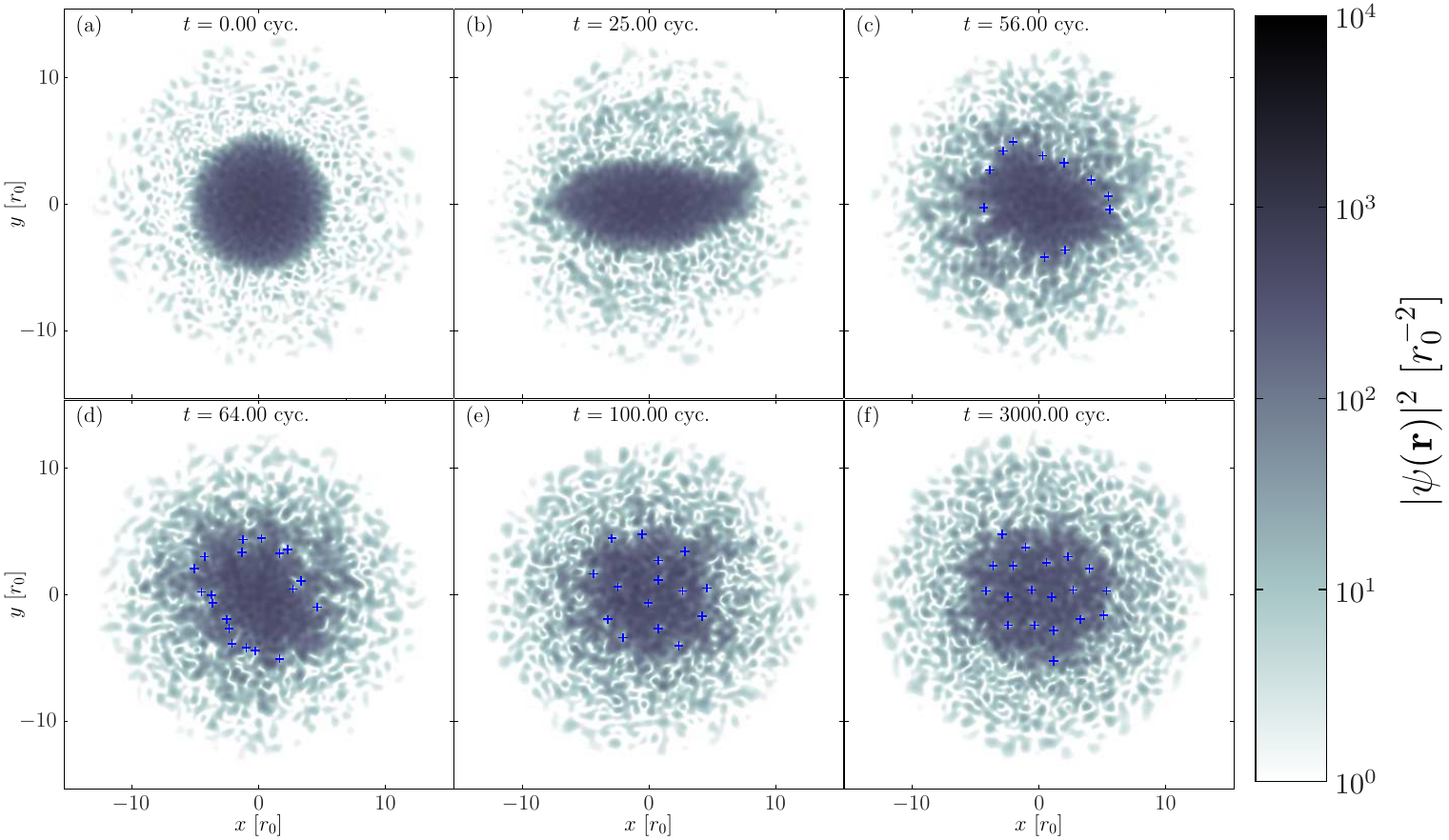}
	\caption{\label{fig:density_plots1} (Color online) Classical field
	density at representative times during the evolution.  Shown are (a)
	initial condition: lab.  frame ground state plus vacuum noise, (b)
	ejection of material during the dynamical instability, (c) nucleation of
	vortices at the interface between the condensate and the non-condensate
	material, (d) penetration of vortices into the condensate bulk, (e)
	non-equilibrium state with rapid vortex motion, and (f) equilibrium
	state.  Vortices are indicated by $+$, and are shown only where the
	surrounding density of the fluid exceeds some threshold value.
	Parameters of the trajectory are given in the text.}
\end{figure*}

We will focus primarily on the case of $\mu_\mathrm{i} = 14\hbar\omega_r$, and
note for comparison that a pure condensate (i.e., ground GP eigenstate) with the
same number of particles rotating at $\Omega = 0.75 \omega_r$ would contain a
lattice of $\approx18$ vortices.  The response to the stirring is illustrated in
the sequence of density distribution plots shown in
Fig.~\ref{fig:density_plots1}, and exhibits the following key features.

\emph{Dynamical instability and formation of thermal cloud.}--- The initial
state (Fig.~\ref{fig:density_plots1}(a)) quickly becomes elongated, with its
long axis oscillating irregularly relative to the major axis of the trap.  The
quadrupole oscillations are dynamically unstable to the stirring perturbation
\cite{Recati01,Sinha01,Parker06}, and since all the Bogoliubov modes of the
condensate are effectively populated due to the representation of quantum noise
in our simulation, some of the fluctuations grow exponentially.  The effect is
clearly seen in Fig.~\ref{fig:density_plots1}(b), where matter streams off the
tips of the elliptically deformed condensate in the direction of the trap
rotation.  Material is then ejected more or less continuously, forming a ring
about the condensate, with the latter oscillating in its motion and shape.  The
ring soon becomes turbulent and diffuse (Fig.~\ref{fig:density_plots1}(c)),
losing coherence and forming a thermal cloud whose characterisation
(Sec.~\ref{subsec:Thermo_params}) is a central theme of this paper.  In
Sec.~\ref{subsubsec:local_rot_prop} we show from an analysis of the angular
momentum and density distribution that by $t=50$ cyc.  the outer part of the
cloud (beyond $r \approx 7r_0$) rotates as a normal fluid in rotational
equilibrium with the drive.

\emph{Vortex nucleation.}--- After the thermal cloud has formed, surface
oscillations of the condensate grow in magnitude
(Fig.~\ref{fig:density_plots1}(d)), in accordance with the prediction of
Williams \emph{et al.}, \cite{Williams2002a} that such oscillations are unstable
in the presence of a thermal cloud.  These fluctuations form a transition region
between the central condensate bulk, characterised by comparatively smooth
density and phase distributions (see also
Fig.~\ref{fig:phase_and_correl_time}(a)), and an outer region of thermally
occupied highly energetic excitations.  Long-lived vortices are nucleated in
this transition region, and then begin to penetrate into the edges of the
condensate region.  Initially the vortices penetrate only small distances
$r\approx r_0$ with rapid and irregular motion, and visibly lag the rotation of
the drive.  Penckwitt \emph{et al.} \cite{Penckwitt02} have previously given a
description of this vortex nucleation and penetration using a model in which a
rotating thermal cloud was included phenomenologically.

\emph{Vortex array formation.}--- The nucleated vortices remain near the edge of
the condensate for a considerable time before they begin to penetrate
substantially into the condensate.  Eventually vortices pass through the central
region (by $t \approx 75$ cyc.  in this simulation), with trajectories that are
largely independent, except when two vortices closely approach each
other~\cite{note4}.  The density of vortices within the central region gradually
increases (Fig.~\ref{fig:density_plots1}(e)), and the vortices begin to form a
rather disorderly assembly, which still lags behind the drive in its overall
rotation about the trap axis.  The erratic motion of individual vortices and
their differential rotation with the respect to drive then gradually slow, and
they become more localised, until by $t \approx 150$ cyc.  they have formed into
an array rotating at the same speed as the drive.  This formation process can be
interpreted as the damping of vortex motion by the mutual friction between them
and the rotating thermal cloud \cite{Sonin87, Donnelly91}.  Alternatively, we
can interpret the motion of the vortices as the result of low energy excitations
of the underlying vortex lattice state \cite{Sonin87}, which are gradually
damped by their interaction with high energy excitations \cite{Fedichev98,
Giorgini98, Pitaevskii97}.  The coupling of the high-energy modes to the
low-energy excitations of the lattice state, drives their distribution towards
equilibrium \cite{Sagdeev88, Sinatra99}.  The high energy modes comprising the
thermal cloud equilibrate to a distribution with identifiable \emph{effective}
thermodynamic parameters (see Sec.~\ref{subsec:Thermo_params}) on a relatively
short timescale ($\sim 100$ cyc.).  While the effective temperature remains
approximately fixed, the effective chemical potential subsequently increases
over a longer timescale ($\sim 3000$ cyc.), as the distribution is shaped by its
interaction with the low energy excitations (including lattice excitations).  By
$t\approx1000$ cyc.  the vortex motion reaches its equilibrium level, and the
equilibrium populations of lattice excitations at this temperature are such that
the vortex distribution does not `crystallize' into a rigid lattice.  The
damping of vortex motion is analysed quantitatively in
Sec.~\ref{subsec:motional_damping}.

\section{Analysis}\label{sec:analysis}

In this section we present the set of measurements we use to analyse the
properties of the thermal cloud, and to identify the remaining condensate
fraction.  We begin by making a simple estimation of the energy and temperature
the thermal cloud will acquire during the lattice formation.  We obtain the
chemical potential of the thermal cloud by using a self consistent fitting
procedure for the atomic distribution function, which also provides a more
accurate measure of the temperature.  We characterise the rotational properties
of different spatial regions of the system, which provides evidence that during
the vortex nucleation stage, the central part behaves as a superfluid, while the
outer cloud behaves as a classical gas, rotating like a rigid body.

The issue of condensate identification is a very important one, and is most
commonly done in terms of spatial correlation functions, using ensemble averages
of quantum mechanical operators \cite{Dalfovo99,Leggett01}.  However, this is
not well suited to the case of vortex lattice formation, due to varying vortex
configurations within the ensemble, with correspondingly destructive effect on
the spatial order of the system.  In the classical field description of a finite
temperature Bose gas, time averages are often substituted for ensemble averages,
making use of the ergodic hypothesis \cite{Blakie05}.  We show, in
Sec.~\ref{subsec:Temporal} that temporal correlation functions can be used to
characterise the local coherence of the field, and consequently determine the
extent of the condensate mode and its eigenvalue.

Following the first appearance of an identifiable vortex array, there is a long
period over which it slowly relaxes to an equilibrium state.  In
Sec.~\ref{subsec:motional_damping} we measure the motional damping of the
vortices and interpret it in terms of their interaction with the thermal
component of the field \cite{Fedichev99}.  Note that all results and figures in
this section (Figs.~\ref{fig:example_temp_fit}-\ref{fig:vortex_damping1})
correspond to the simulation shown in Fig.~\ref{fig:density_plots1}.

\subsection{Thermodynamic parameters}\label{subsec:Thermo_params}
\subsubsection{Analytic predictions}

Since our system conserves both normalisation and energy in the rotating frame,
some simple analytic predictions can be made about the development of the
thermal component of the field, using the Thomas-Fermi (TF) approximation.  In a
frame rotating at angular velocity $\Omega$, the ground state is a vortex
lattice which, neglecting the effect of the vortex cores, has Thomas-Fermi
wavefunction \begin{equation} \psi_\mathrm{TF}^{\Omega}=\sqrt{\frac{\mu_\Omega
-m\left(\omega_r^2-\Omega^{2}\right) r^{2}/2}{U_\mathrm{2D}}}\,\Theta
\left(\mu_\Omega-m\left(\omega_r^2-\Omega^{2}\right) \frac{r^{2}}{2}\right),
\label{eq:psiTF}\end{equation}
 where $ $$\mu_\Omega$ is the chemical potential of the rotating frame solution.
From Eq.~(\ref{eq:psiTF}) we obtain the number of particles $
$$N_\mathrm{TF}^\Omega$ and energy $E_\mathrm{TF}^\Omega$ for the lattice state
\begin{eqnarray} N_\mathrm{TF}^\Omega
&=&\frac{\pi\mu_\Omega^{2}}{U_\mathrm{2D}m\left(\omega_r^2-\Omega^{2}\right)},
\label{eq:NTF_vortex}\\ E_\mathrm{TF}^\Omega &=&\frac{2}{3}\mu_\Omega
N_\mathrm{TF}^\Omega.\label{eq:ETF_vortex} \end{eqnarray}

The corresponding quantities for the vortex free inertial frame ground state are
given by setting $\Omega=0$.  We note that $\mu_0 \equiv \mu_{\Omega=0}$ has the
same value in both the stationary and rotating frames (since this solution is
non-rotational).  If we now compare a lattice state and a vortex free state,
both fully condensed and with equal occupation, we see from
Eq.~(\ref{eq:NTF_vortex}) that the chemical potentials are related by
\begin{equation}\label{eq:reduced_mu}
	\mu_\Omega=\mu_0\sqrt{1-\Omega^{2}/\omega_r^2},
\end{equation} and hence from Eq.~(\ref{eq:ETF_vortex}) that their rotating
frame energies are related by \begin{equation}
E_\mathrm{TF}^\Omega=E_\mathrm{TF}^0\sqrt{1-\Omega^{2}/\omega_r^2}.
\label{eq:Ecompare} \end{equation} The excess energy of the vortex free ground
state over the lattice state of the same number of atoms, \begin{equation}
\Delta E\equiv
E_\mathrm{TF}^0-E_\mathrm{TF}^\Omega\label{eq:delta_E}\end{equation} can be
significant, and for example is approximately $E_\mathrm{TF}^0/3$ for an angular
velocity of $\Omega=0.75\omega_r$.  Thus in our stirring scenario, the rotating
equilibrium state reached at the end of the process must have less atoms and
less energy than the initial vortex free state, and the excess energy and atoms
constitute a thermal cloud \cite{note5}.  The exact result for the excess energy
of the classical field, obtained from the simulations, will depend upon the
depletion of the condensate mode required to form the thermal cloud and the
mutual interaction of the cloud with the condensate.  For small thermal
fractions this energy could be found by means of a calculation similar to that
described in Ref.~\cite{Sinatra00}.  In the simplest approximation we assume the
limit of a small thermal fraction, in which the energies of the condensate and
thermal field are additive, with Eq.~(\ref{eq:delta_E}) approximating the excess
energy the thermal field contains.  We further assume that in equilibrium this
energy will be classically equipartitioned over weakly interacting harmonic
oscillators, in the spirit of the Bogoliubov approximation \cite{Lobo04}.  The
equilibrium temperature can thus be predicted by
\begin{equation}\label{eq:temperature_prediction} T=\frac{\Delta
E}{\mathcal{M}-1}, \end{equation}
 with $\mathcal{M}$ the condensate band multiplicity.  For a given
initial chemical potential we see therefore that the equilibrium temperature
will be strongly dependent on the basis size.  However, we shall see in
Section~\ref{sec:parameters}, Eq.~(\ref{eq:reduced_mu}) provides a very good
estimate of the final chemical potential reached by the field, which is
essentially independent of cutoff.

\subsubsection{Self-consistent fitting}\label{subsec:fitting} 
\begin{figure}
	\includegraphics[width=0.45\textwidth]{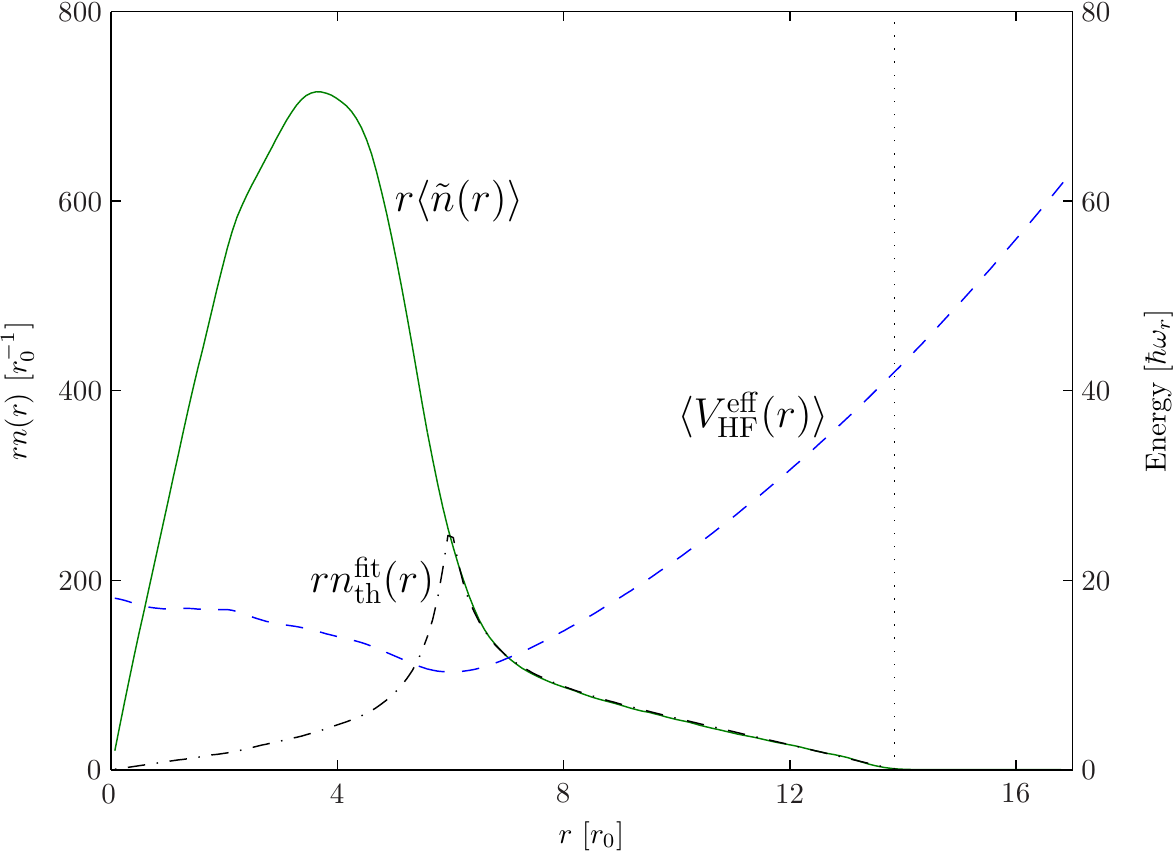}
	\caption{\label{fig:example_temp_fit} (Color online) Fitting procedure
	for the thermodynamic parameters.  Shown are: the (time-averaged) radial
	density distribution times radius (solid line), the effective potential
	experienced by non-condensate atoms times radius (dashed line), and the
	fitted distribution of non-condensed atoms (dash-dot line).  The dotted
	line indicates the classical turning point corresponding to energy
	cutoff $E_\mathrm{R}$.  The data shown corresponds to the period
	$t=3000-3010$ cyc.  (Fig.~\ref{fig:density_plots1}(f)).}
\end{figure}

We expect a
procedure such as the self-consistent Hartree-Fock approximation described in
\cite{Giorgini97} to provide a good estimate for both $\mu$ and $T$.  However in
order to avoid performing the iterative procedure presented in \cite{Giorgini97}
we use a simpler approximation to this description, similar to that presented in
\cite{Naraschewski98}, but with some necessary modifications.  The TF
approximation for the condensate mode employed in \cite{Naraschewski98} is a
poor choice for the situation considered here, where the distribution of
vortices is irregular and the density of vortices is low.  Secondly a
straightforward application of the model of \cite{Naraschewski98} to our system
leads to divergences, as the semiclassical density distribution of non-condensed
atoms becomes \begin{equation}\label{eq:Naraschewski_ntherm}
	n_\mathrm{th}(\mathbf{x}) = \frac{1}{\lambda_T^2}
	g_1[e^{-(V_\mathrm{eff}(\mathbf{x}) + 2U_\mathrm{2D}
	(n_\mathrm{c}(\mathbf{x}) + n_\mathrm{th}(\mathbf{x})) - \mu)/kT}],
\end{equation} with $V_\mathrm{eff}(\mathbf{x})=m(\omega_r^2-\Omega^2)r^2/2$ the
centrifugally dilated trapping potential in the rotating frame.  The Bose
function $g_1(z)$ diverges logarithmically \cite{Pathria96} as $z\rightarrow1$,
and thus the function $n_\mathrm{th}(\mathbf{x})$ diverges where
$V_\mathrm{eff}(\mathbf{x}) +2U_\mathrm{2D} (n_\mathrm{c}(\mathbf{x}) +
n_\mathrm{th}(\mathbf{x})) \rightarrow \mu$.  Neglecting the mutual mean-field
repulsion $2U_\mathrm{2D} n_\mathrm{th}(\mathbf{x})$ of non-condensed atoms (as
in \cite{Naraschewski98}) would therefore leave us with divergences in
$n_\mathrm{th}(\mathbf{x})$.  We are thus led to develop a fitting function for
the non-condensed density $n_\mathrm{th}(\mathbf{x})$ only, and which therefore
depends only on the \emph{total} field density $n(\mathbf{x})$, which we measure
from the classical field trajectory.  The inclusion of the mean-field
interaction between non-condensed atoms ensures that our function remains
regular so long as the (fitted) chemical potential is less than the minimum
value of the effective potential experienced by non-condensed atoms, a
restriction which is not expected to constrain our fit in an unphysical manner.
In this way we avoid both having to distinguish between condensed and
non-condensed material in forming the mean-field potential, and having to
iterate a self-consistent model to convergence.  Our approach to determining
$\mu$ and $T$ is as follows: The one-body Wigner function corresponding to a
Bose field $\hat{\phi}(\mathbf{x})$ is defined \begin{equation}
	F(\mathbf{x},\mathbf{k}) \equiv \int d\mathbf{y} \langle
	\hat{\phi}^\dagger(\mathbf{x}+\frac{\mathbf{y}}{2})\hat{\phi}
	(\mathbf{x}-\frac{\mathbf{y}}{2})\rangle e^{i\mathbf{k}\cdot\mathbf{y}},
\end{equation} and we assume here the non-condensate atoms in the condensate
band to be well described by such a distribution, for which we further assume
the approximate semiclassical form \cite{Castin01,Bagnato87}
\begin{equation}\label{eq:semiclassical_Wigner_dist}
	F(\mathbf{x},\mathbf{k}) =
	\frac{1}{\exp[(\epsilon(\mathbf{x},\mathbf{k}) - \mu)/T] - 1}.
\end{equation} The semiclassical energy is \begin{equation}
	\epsilon(\mathbf{x},\mathbf{K}) = \frac{\hbar^2\mathbf{K}^2}{2m} +
	V_\mathrm{eff}^\mathrm{HF}(\mathbf{x}),
\end{equation} with $\mathbf{K}$ the wave vector adjusted to the rotating frame
\cite{Bradley08}, and the Hartree-Fock effective potential in the rotating frame
is \begin{equation}
	V^{\mathrm{HF}}_\mathrm{eff}(\mathbf{x}) = m(\omega_r^2 -
	\Omega^2)\frac{r^2}{2} + 2U_\mathrm{2D} n(\mathbf{x}),
\end{equation} with the total field density the sum of the condensate and
non-condensate (thermal) contributions $n(\mathbf{x})=n_\mathrm{c}(\mathbf{x}) +
n_\mathrm{th}(\mathbf{x})$.  The factor of two here is the strength of the
interaction between pairs of atoms when at least one of the pair is
non-condensed \cite{Giorgini97,Naraschewski98,Castin01}.  The classical field
limit of Eq.~(\ref{eq:semiclassical_Wigner_dist}) is obtained when
$F(\mathbf{x},\mathbf{k}) \gg 1$, which occurs when the parameter
$(\epsilon(\mathbf{x},\mathbf{k}) -\mu)/kT \ll 1$, and is given by its first
order expansion in this parameter,
\begin{equation}\label{eq:classical_limit_dist}
	F_\mathrm{c}(\mathbf{x},\mathbf{k}) =
	\frac{kT}{\epsilon(\mathbf{x},\mathbf{k}) - \mu}.
\end{equation} It is important to note that although
Eq.~(\ref{eq:classical_limit_dist}) is obtained here as the high occupation
approximation to the bosonic distribution
Eq.~(\ref{eq:semiclassical_Wigner_dist}), (describing the classical
equipartitioning of energy over phase-space cells of volume $1/h^2$), we expect
it to apply to the equilibrium state of a weakly interacting classical field
system even if the level occupations are $\lesssim 1$
\cite{Sinatra01,Lobo04,note6}.  With the \emph{local} rotating frame wave vector
cutoff defined by $\hbar^2K_\mathrm{R}(\mathbf{x})^2/2m\equiv \max\{E_\mathrm{R}
- V_\mathrm{eff}^\mathrm{HF}(\mathbf{x}),0\}$ \cite{Bradley08}, we find for the
spatial distribution \begin{eqnarray}\label{eq:fit_function1}
	n_\mathrm{th}^\mathrm{fit}(\mathbf{x},\mu,T) &=&
	\int_0^{K_\mathrm{R}(\mathbf{x})} \frac{K\;dK}{2\pi} F_c(\mathbf{x},K)
	\\ \nonumber
		     &=& \frac{1}{\lambda_\mathrm{dB}^2}
		     \ln\left[\frac{E_\mathrm{R} -
		     \mu}{V_\mathrm{eff}^\mathrm{HF}(\mathbf{x}) - \mu}\right],
\end{eqnarray} on $|\mathbf{x}| \leq r_\mathrm{tp}$, with $r_\mathrm{tp}=\sqrt{2
E_\mathrm{R}/m(\omega_r^2-\Omega^2)}$ the semiclassical turning point of the
condensate band in the rotating frame, and
$\lambda_\mathrm{dB}=\sqrt{2\pi\hbar^2/m k_\mathrm{B}T}$ the thermal de Broglie
wavelength.  This expression constitutes an appropriate fitting function for the
non-condensed component of the classical field, which requires no explicit
knowledge of the distribution of condensed atoms \cite{note7}.  To obtain an
estimate of $\mu$ and $T$ for our system, we sample the azimuthally averaged
field density $\tilde{n}(r)\equiv\frac{1}{2\pi}\int_0^{2\pi}d\theta \;
n(r,\theta)$ at $N_t$ equally spaced times over some period
$t\in[t_0-T/2,t_0+T/2]$, and average over the sample times to obtain $\langle
\tilde{n}(r)\rangle \equiv \sum_{i=1}^{N_t} \tilde{n}(r,t_i)/N_t$.  From this we
construct the effective potential

\begin{equation}\label{eq:eff_potl}
	\langle V_\mathrm{HF}^\mathrm{eff}(r) \rangle =
	m(\omega_r^2-\Omega^2)r^2/2 + 2U_\mathrm{2D}\langle\tilde{n}(r)\rangle.
\end{equation} We then take as our fitting function
\begin{equation}\label{eq:fit_function2}
	n_\mathrm{th}^\mathrm{fit}(r;\mu,T) = \frac{1}{\lambda_\mathrm{dB}^2}
	\ln \left[\frac{E_\mathrm{R} - \mu}{\langle
	V^\mathrm{HF}_\mathrm{eff}(r) \rangle - \mu}\right].
\end{equation} As our fitting function represents only the non-condensed
component our field, we are obliged to avoid including significant condensate
density in our fit to $\langle \tilde{n}(r)\rangle$.  We therefore perform our
fit of $n_\mathrm{th}^\mathrm{fit}(r)$ to $\langle \tilde{n}(r)\rangle$ over the
domain $r\in[r_-,r_\mathrm{c}]$, with $r_-$ the location of the occurrence of
the minimum value of the effective potential, i.e., $\langle
V^\mathrm{HF}_\mathrm{eff}(r_-)\rangle = \min\{\langle
V^\mathrm{HF}_\mathrm{eff}(r)\rangle\}$, which should approximately mark the
peak of the non-condensed fraction and thus the boundary of the condensate
\cite{Naraschewski98}.  Within the central region $r<r_-$, the fitted density
$n_\mathrm{th}^\mathrm{fit}(r)$ thus decays monotonically as $r\rightarrow0$.
An example of such a fit is shown in Fig.~\ref{fig:example_temp_fit}.

\begin{figure}
	\includegraphics[width=0.45\textwidth]{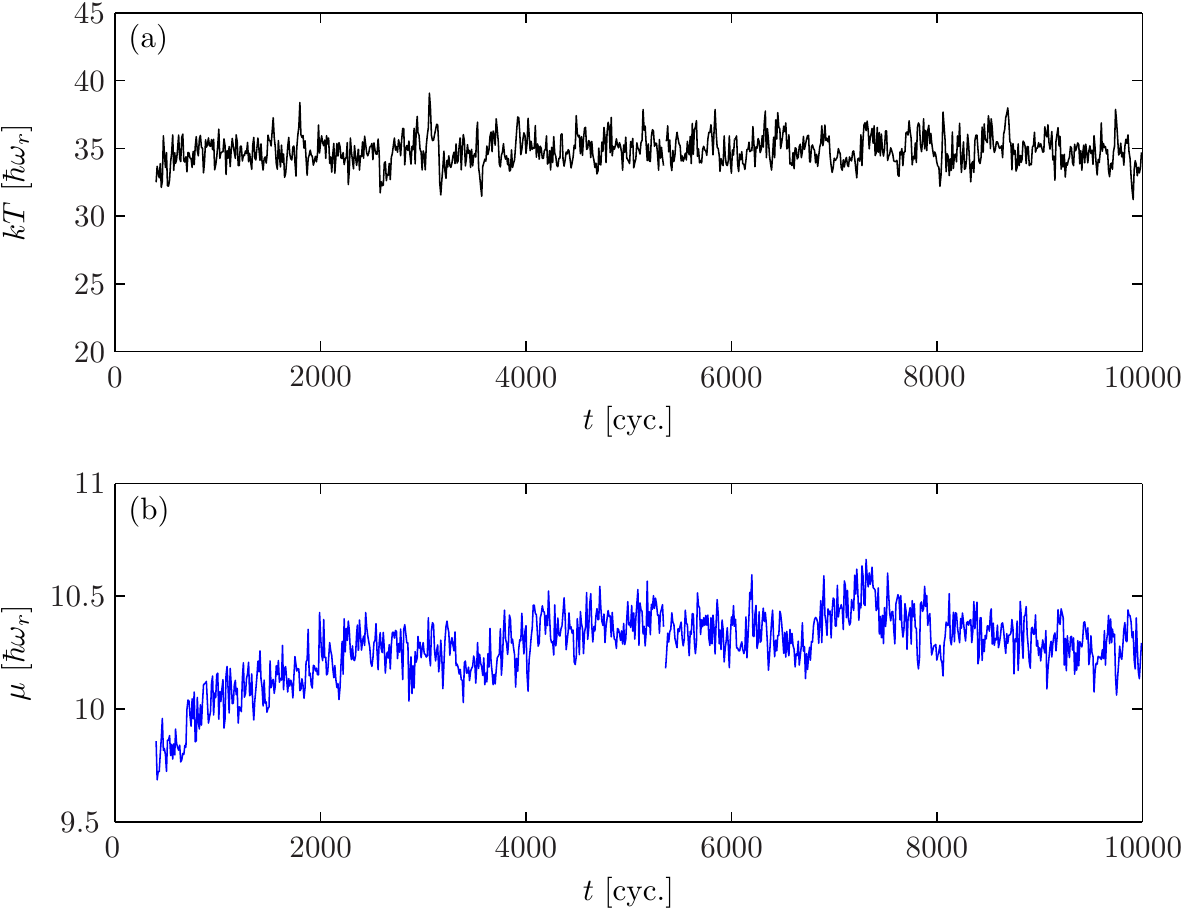}
	\caption{\label{fig:T_mu_from_fit} (Color online) Evolution of the
	effective (a) temperature and (b) chemical potential of the
	non-condensed atoms, as measured by the semiclassical fitting
	procedure.}
\end{figure}

After the initial strongly non-equilibrium dynamics following the dynamical
instability (i.e.  by $t\approx400$ cyc.), the averaged field densities $\langle
\tilde{n}(r)\rangle$ are fit well by the function given in
Eq.~(\ref{eq:fit_function2}), and we conclude that the higher energy components
of the field have reached a quasistatic equilibrium.  We follow thereafter the
evolution of the temperature and chemical potential with time, which we present
in Fig.~\ref{fig:T_mu_from_fit}.  The temperature
(Fig.~\ref{fig:T_mu_from_fit}(a)) appears to be essentially constant within the
accuracy of the measurement procedure, while the chemical potential
(Fig.~\ref{fig:T_mu_from_fit}(b)) shows a definite upward trend over the course
of the field's evolution.  In Sec.~\ref{subsec:Frequency_dist} we compare the
chemical potential of the cloud determined from our fitting procedure, with the
chemical potential of the condensate, which we extract by considering the
temporal frequency components present in the field.

\subsection{Rotational parameters}\label{subsec:Rotational_params} A
characteristic feature of superfluids is that they resist any attempt to impart
a rotation to them, and they acquire angular momentum through the nucleation of
vortices \cite{note8}, which endow the superfluid with \emph{quantized} flow
circulation (see e.g.  \cite{Donnelly91}).  Thus while a classical fluid in
equilibrium with a container rotating at angular velocity $\Omega$ satisfies the
equality

\begin{equation}
	\langle L_z\rangle = \Omega I_\mathrm{c},
\end{equation} where the classical moment of inertia is defined
\begin{equation}\label{eq:classical_inertia}
	I_\mathrm{c} \equiv m\langle r^2\rangle,
\end{equation} a superfluid in equilibrium with such a container will in general
fail to do so.  That is to say, the moment of inertia per particle
\begin{equation}\label{eq:quantum_inertia}
	I\equiv \frac{\langle L_z\rangle}{N\Omega},
\end{equation} will not equal the classical value $I_\mathrm{c}/N$ in such a
state.  We note that for pure superfluids, such as a dilute Bose gas at $T=0$
(which is well described as a ground state of the GP equation), the steady state
angular momentum is typically much less than the classical value determined by
its mass distribution \cite{Zambelli01}.  As the rotation frequency $\Omega$ of
the GP state increases, the (areal) density of vortices increases
\cite{Garcia-Ripoll01,Feder01}, and is given to leading order by the Feynman
relation \cite{Donnelly91} \begin{equation}\label{eq:Feynman_relation}
	\rho_\mathrm{v}^\mathrm{c} = \frac{m\Omega}{\pi\hbar},
\end{equation} with corrections arising due to the lattice inhomogeneity and the
discrete nature of the vortex array \cite{Sheehy04}.  As the rotation frequency
and thus the vortex density increase, both the classical and quantum moments of
inertia increase also, and the difference between the two moments vanishes in
the limit of rapid rotations $\Omega \rightarrow \omega_r$ \cite{Feder01}.

\subsubsection{Vortex density} 
\begin{figure}
	\includegraphics[width=0.45\textwidth]{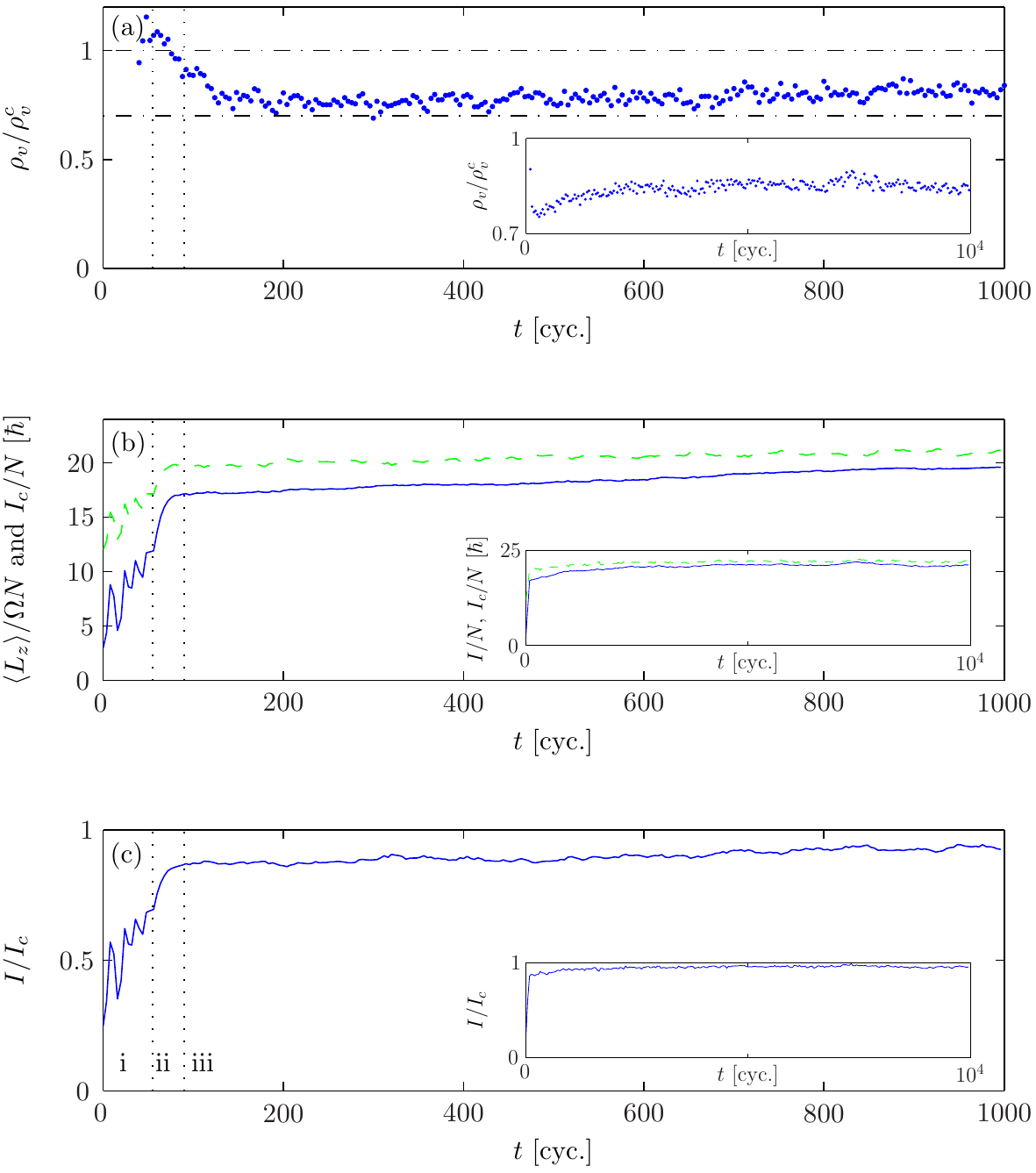}
	\caption{\label{fig:angmom_etc1} (Color online) Rotational response of
	the field.  Shown are: (a) Vortex density, with dash-dot lines
	indicating $\rho_\mathrm{v}/\rho_\mathrm{v}^\mathrm{c} = 0.7$ and
	$\rho_\mathrm{v}/\rho_\mathrm{v}^\mathrm{c} =1$ for reference, (b)
	classical (dashed line) and quantum moments of inertia, and (c) ratio of
	quantum and classical moments of inertia.  Insets show the behaviour of
	the same quantities over a longer timescale.  Note that the initial
	$L_z$ is finite due to the initial vacuum occupation.  For reference
	dotted lines in (a-c) separate phases of the evolution discussed in the
	text.}
\end{figure}

To monitor the increase in vortex density during
the system evolution, we count the vortices of positive rotation sense occurring
within a circular counting region of radius equal to the TF radius of the
initial state, $R_\mathrm{TF} = \sqrt{2\mu_\mathrm{i}/m\omega_r^2}$.  Due to the
rotational dilation of the condensate in the rotating frame this region lies
within the central bulk of the condensate, and in this manner we attempt to
avoid including in the count the short-lived vortices constantly created and
annihilated at the condensate boundary.  At equally spaced times over a period
of 4 trap cycles about time $t_i$ we obtain the average number of vortices in
the counting radius, $\langle n_\mathrm{v}(t_i) \rangle$, and from this we
determine the average vortex density in the counting region
\begin{equation}\label{eq:vortex_density}
	\rho_\mathrm{v}(t_i) = \frac{\langle n_\mathrm{v}(t_i) \rangle}{\pi
	R_\mathrm{TF}^2}.
\end{equation} In Fig.~\ref{fig:angmom_etc1}(a) we plot
Eq.~(\ref{eq:vortex_density}) as a fraction of the prediction of
Eq.~(\ref{eq:Feynman_relation}).  During the dynamical instability and initial
nucleation of vortices the vortex densities measured in this manner are
spuriously high, due to the counting of phase defects in the turbulent
non-equilibrium fluid rather than long-lived vortices in the condensate bulk.
After this initial period, i.e.  from $t\approx 150$ cyc.  onwards, we see a
gradual increase in $\rho_\mathrm{v}$ as the condensate slowly relaxes and
admits more vortices into its interior, and approaches rotational equilibrium
with the trap.  By $t\approx 6000$ cyc., the density appears to have essentially
saturated at a level $\rho_\mathrm{v} \approx 0.85\rho_\mathrm{v}^\mathrm{c}$.
Such a value seems reasonable for the rotation rate and condensate size
considered here, which are such that inhomogeneity and discreteness effects will
be significant \cite{Sheehy04}.  Moreover, it seems plausible that the high
degree of lattice excitation here increases the energetic cost of vortices above
that assumed in the mean-field description of \cite{Sheehy04}.

\subsubsection{Local rotational properties}\label{subsubsec:local_rot_prop} 
\begin{figure}
	\includegraphics[width=0.45\textwidth]{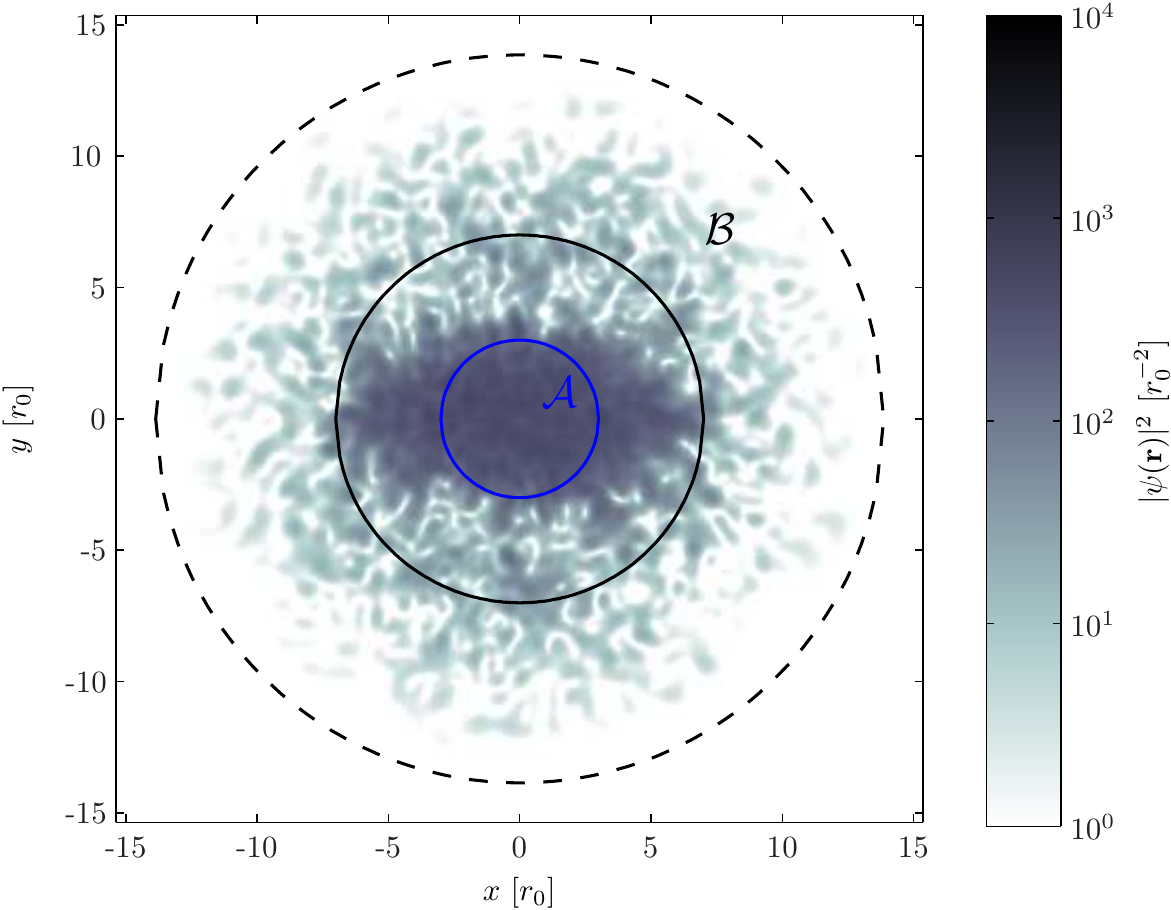}
	\caption{\label{fig:angmom_annulus} (Color online) Regions used for the
	evaluation of angular momenta and moments of inertia.  Disc
	$\mathcal{A}$ lies entirely within the condensate, where the flow is
	(initially) irrotational.  Annulus $\mathcal{B}$ (between the solid
	black line and the dashed line indicating the classical turning point of
	the condensate band) contains only non-condensed, normal fluid.  The
	field density shown is that at $t=50$ cyc.}
\end{figure}

In
order to characterise the rotation of the outer cloud formed from ejected
material as well as that of the central bulk of the field we measure
\emph{localised} expectation values, i.e.  expectation values over a restricted
spatial region.  In this way we can compare the mass distribution and angular
momentum of particular regions, in order to characterise the localised
properties of the field.  We define the expectation value of an operator
$\mathcal{O}$ on spatial domain $\mathcal{D}$ by
\begin{equation}\label{eq:restricted_expvalue}
	\langle \mathcal{O} \rangle_\mathcal{D} = \int_\mathcal{D} d\mathbf{x}\;
	\psi^* \mathcal{O} \psi,
\end{equation} where the $\mathcal{D}$ will be either region $\mathcal{A}$ or
$\mathcal{B}$ illustrated in Fig.~\ref{fig:angmom_annulus}.  Focussing our
attention on the outer annulus $\mathcal{B}$, we calculate the classical moment
of inertia of material in the annulus \begin{equation}
	(I_\mathrm{c})_\mathcal{B} = m\langle r^2\rangle_\mathcal{B},
\end{equation} and the angular momentum of this material $\langle L_z
\rangle_\mathcal{B}$.  From measurements over the period $t=50-51$ cyc.  we find
that \begin{equation}
	\frac{\langle L_z\rangle_\mathcal{B}}{\Omega} =
	(I_\mathrm{c})_\mathcal{B},
\end{equation} to within 4\%, indicating that the cloud in this region is
rotating as a normal fluid in rotational equilibrium with the drive.  Averaging
samples of these expectation values over a period of a trap cycle, we find for
the time averages $\{\langle \tilde{L_z} \rangle_\mathcal{B} / \Omega \}= 1.025
\{(\tilde{I_\mathrm{c}})_\mathcal{B}\}$.

Turning our attention to the central disc $\mathcal{A}$, we find that over the
same period, $\{\langle \tilde{L_z}\rangle_\mathcal{B}\}/\Omega
=0.080\{(I_\mathrm{c})_\mathcal{B}\}$.  The central condensate bulk at this time
thus remains approximately stable against vortex nucleation, but possesses some
small angular momentum due to its shape oscillations, in contrast to the ejected
material which has lost its superfluid character and has come to rotational
equilibrium with the drive.

\subsubsection{Global rotational properties} We consider now the rotational
properties of the entire field, by evaluating expectation values as above over
the full $xy$-plane, (i.e.  in Eq.~(\ref{eq:restricted_expvalue}) we set
$\mathcal{D}\rightarrow\mathbb{R}^2$).  A measure of the field's rotation is
given by the angular momentum $L_z$.  We plot the ratio $I/I_\mathrm{c}$ to
characterise the response of the atomic field to the imposed rotation, as both
the field's angular momentum and its mass distribution change with time.  In
Fig.~\ref{fig:angmom_etc1}(b) we plot the evolution of these quantities as time
evolves and in Fig.~\ref{fig:angmom_etc1}(c) we plot the evolution of their
ratio.

Comparing the evolution of these quantities and that of the field's density
distribution (c.f.  Fig.~\ref{fig:density_plots1}), we identify three roughly
distinct phases of the system evolution: (i) Excitation of unstable surface mode
oscillations accompanied by permanent increase in angular momentum (c.f.
Fig.~\ref{fig:density_plots1}(b)), (ii) further increase in angular momentum and
$I/I_\mathrm{c}$ as vortices are nucleated into the condensate bulk
(Fig.~\ref{fig:density_plots1}(c-e)), and (iii) gradual approach of the field to
rotational equilibrium with the drive (Fig.~\ref{fig:density_plots1}(f)).  Note
that although the finite temperature equilibria here contain a significant
normal fluid component, the quantum moment of inertia is still suppressed below
the classical value due to the presence of the superfluid component.

\subsection{Temporal analysis}\label{subsec:Temporal} \subsubsection{Issues of
condensate identification} A central concern in classical field simulations is
the identification of the condensate mode and its occupation.  In situations
such as \cite{Norrie05, Norrie06a, Norrie06b} where the collisional dynamics
dominate, and dynamics of interest occur on a short timescale, the condensate
mode can be identified within a $U(1)$ symmetry-broken description as the
\emph{coherent} fraction of the field \begin{equation}
	\psi_\mathrm{coh} = \langle \hat{\psi} \rangle,
\end{equation} where the expectation value in the many-body state $\langle
\cdots \rangle$ can be formally related to moments of the classical field
\cite{Gardiner00, Gardiner02}.

However, on longer timescales the classical field is expected to evolve to a
thermal distribution \cite{Sinatra02,Davis01}, and so we abandon the formal
operator correspondences and examine the \emph{classical} statistics of the
trajectory ensemble directly.  Furthermore, the chaotic nature of the system
means that trajectories which are initially close in phase-space will, in
general, diverge as it evolves.  Thus even in an idealised situation in which
distinct trajectories each contain a highly occupied mode undergoing coherent
phase rotation, the phase of this mode at a particular time $t$ will vary
between trajectories, and the identification of the condensate mode with the
field operator mean yields a condensate fraction which decays spuriously as the
neighbouring trajectories dephase.  This is a natural consequence of the
symmetry-broken condensate definition, and occurs even for small non-condensate
fractions \cite{Lewenstein96,Castin98}, however the dephasing is intuitively
expected to occur more rapidly in systems with a greater incoherent fraction.

However, global gauge symmetry breaking has no effect on the one body density
matrix \begin{equation}
	\rho(\mathbf{x},\mathbf{x}') \equiv \langle
	\hat{\psi}^\dagger(\mathbf{x})\hat{\psi}(\mathbf{x}') \rangle,
\end{equation} or its classical field analogue \cite{Blakie05}.  In a strongly
thermal system, the ergodic hypothesis may be employed to approximate such an
ensemble expectation value by a time average.  Blakie and coworkers
\cite{Blakie05} have used the Penrose-Onsager \cite{Penrose56} definition of the
condensate mode (the single eigenmode of $\rho(\mathbf{x},\mathbf{x}')$ with an
occupation comparable to the total (mean) number of particles), to extract a
condensate fraction from individual classical field trajectories, in equilibrium
and quasi-equilibrium scenarios.

However, the Penrose-Onsager criterion is not appropriate for the current
situation, which exhibits strongly non-equilibrium behaviour.  As noted in
\cite{Leggett01}, in such situations the Penrose-Onsager definition may fail to
correctly describe the amount of condensation in the system.  Further complexity
is added by the presence of a non-trivial phase structure such as that of a
condensate containing an array of vortices.  Indeed even at $T=0$ in an axially
symmetric trap, the rotating frame ground state is in general highly degenerate
due to the spontaneous breaking of the rotational symmetry of the many-body
Hamiltonian \cite{Seiringer08}.  The corresponding one-body density matrix thus
fails to describe a single highly occupied mode.  While in our case the axial
symmetry of the Hamiltonian is \emph{explicitly} broken by the potential
anisotropy $V_\epsilon(\mathbf{x})$, distinct configurations of vortices
distinguish states of the system in a way which cannot be cancelled by global
phase rotations.  One-body density matrices calculated from either ensemble or
ergodic time averages thus describe a non-simple or \emph{fragmented} condensate
state (see \cite{Leggett01}).  As this behaviour might be expected of the true
one-body density matrix in such a situation, we regard this to be a lack in
generality of the Penrose-Onsager criterion itself.  A quantification of the
condensate population or mode at a single time would therefore require a
description in terms of higher order correlation functions.

\subsubsection{Frequency distribution}\label{subsec:Frequency_dist} Previous
authors studying equilibrium systems have used known quasiparticle bases to
extract frequency distributions from classical field simulations
\cite{Davis02,Brewczyk04}.  In a general non-equilibrium situation, with
significant non-condensate populations, suitable modes can not be defined.  We
therefore use an alternative strategy, and extract information about the
distribution of modes in the trap by analysing the frequencies present at
particular points in position space.  We define the power spectrum of the
classical field $\psi$ at position $\mathbf{x}$ about time $t_0$

\begin{equation}
	H(\mathbf{x},\omega;t_0) =
	|\mathfrak{F}_{t_0}^T\{\psi(\mathbf{x},t)\}|^2,
\end{equation} where $\mathfrak{F}_{t_0}^T\{f(t)\}$ denotes the Fourier
coefficient \begin{equation}
	\mathfrak{F}_{t_0}^T\{f(t)\} \equiv \frac{1}{T}\int_{t_0-T/2}^{t_0+T/2}
	f(t) e^{-i\omega t}dt.
\end{equation} We choose a sampling period of 10 trap cycles so as to be long
compared to the timescales characterising the phase evolution of the condensate
($\tau_\mathrm{p} \sim \hbar/\mu$), while being short compared to the that of
the relaxation of the field ($\tau_R \sim 1000$ cyc.) \cite{Bradley08}, and
choose the sampling interval so as to resolve all frequencies present due to the
combined single particle and mean-field evolutions ($\hbar\omega_\mathrm{max} =
E_\mathrm{R} + 2\mu_\mathrm{i}$).  We average this result over azimuthal angle
$\theta_j$, to form $\tilde{H}(r,\omega,t_0)$.  In order to gauge the relative
strength of the various frequency components at different radii within the
classical region, we normalise the resulting data so that the total power at
each radius $r_j$ is the same.

\begin{figure*}
	\includegraphics[width=0.9\textwidth]{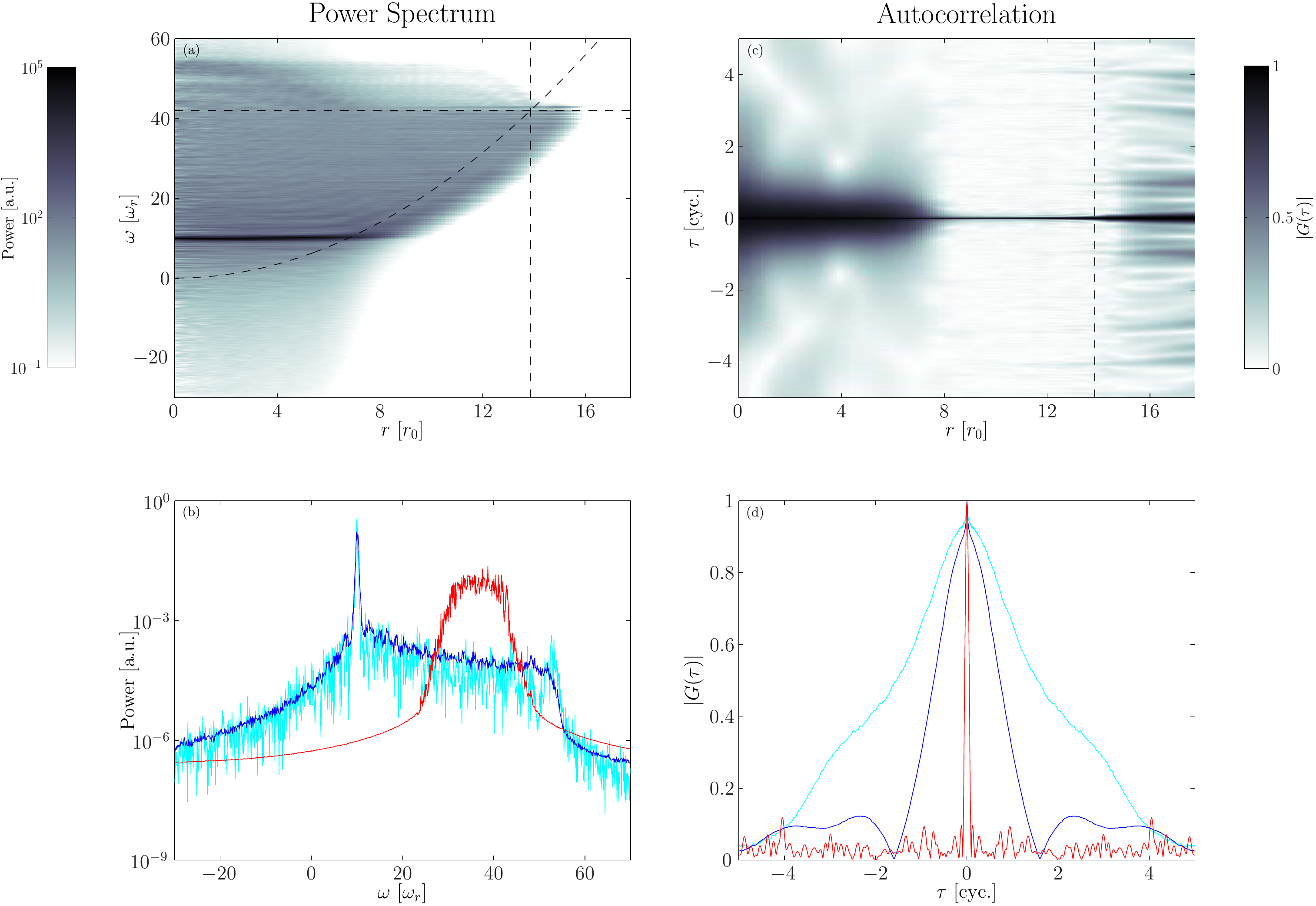}
	\caption{\label{fig:power_and_auto1}(Color online) (a) Azimuthally
	averaged power spectral density (PSD).  Dashed lines indicate the cutoff
	energy, corresponding classical turning point and (centrifugally
	dilated) trapping potential.  (b) PSD traces at particular radii.  The
	black (blue) and dark grey (red) lines correspond to radii $r=3.1894r_0$
	and $r=11.9575r_0$ respectively.  The light grey (cyan) line corresponds
	to the smallest measured radius.  (c) Autocorrelation function (absolute
	value) obtained from the power spectrum.  The dashed line indicates the
	classical turning point of the condensate band.  (d) Autocorrelation
	magnitude at radii corresponding to Fig.~\ref{fig:power_and_auto1}(b).
	Data corresponds to the period $t=9900-9910$ cyc.}
\end{figure*}

In Fig.~\ref{fig:power_and_auto1}(a), we plot the result of such a calculation
over the time interval $t=9900-9910$ trap cycles.  For comparison we have
overlaid the centrifugally dilated potential $V_c(r) =
m(\omega_r^2-\Omega^2)r^2/2$ that would be seen by a classical particle in the
rotating frame, the cutoff frequency $\omega_\mathrm{R} = E_\mathrm{R}/\hbar$,
and the semiclassical turning point $r_\mathrm{tp}$.  Several notable features
are present.  At small radii we see a prominent peak in the power spectrum at
$\omega \approx 10\omega_r$, which we interpret as the coherent phase rotation
frequency of the underlying condensate mode, i.e., the condensate eigenvalue
$\mu_\mathrm{c}/\hbar$.  In Fig.~\ref{fig:power_and_auto1}(b) where the power
spectrum is plotted at 3 particular radii, the peak in the spectrum obtained
close to the trap centre (light gray/cyan line) can be seen clearly.  In
addition we see that frequencies above and below this are present, and we
identify these as due to the thermal occupation of the quasiparticle `particle'
and `hole' modes respectively.  At this same radial value ($r=0.0661r_0$) a
secondary peak is visible at
$\omega\approx52\omega_r=(\mu_\mathrm{c}+E_\mathrm{R})/\hbar$.  An examination
of the data reveals that a smaller peak is also present at
$\omega\approx-32\omega_r=(\mu_\mathrm{c}-E_\mathrm{R})/\hbar$.  This is an
artifact of the projected method which occurs as follows: The $\mathcal{M}$
Bogoliubov modes which diagonalize the Bogoliubov Hamiltonian in the condensate
band in the presence of a stationary condensate mode (GP eigensolution) can be
considered as variational approximations to the lowest-lying `true' Bogoliubov
modes (obtained as the cutoff $E_\mathrm{R}\rightarrow\infty$).  The highest
energy modes in the restricted quasiparticle basis are poor representations of
the true modes, and, to maintain orthogonality of the set, are spuriously
localised to the trap centre, and therefore have spuriously high energies
$\epsilon_k + \mu_\mathrm{c} \approx E_\mathrm{R} +\mu_\mathrm{c}$ due to the
large mean-field interaction there.  Therefore, assuming this behaviour of
excitations to hold in a finite temperature equilibrium of the classical field,
a peak is expected to occur at these frequencies at the trap centre, where the
density of these modes is comparatively large.  However the population contained
in the peak observed here is $\sim0.5\%$ of the total thermal population and as
such we do not expect it to qualitatively affect the relaxation process
\cite{note9}.  The black (blue) line in Fig.~\ref{fig:power_and_auto1}(b)
indicates the power spectrum at a larger radius where the peak is slightly less
prominent.  At larger radii still ($r\gtrsim 9r_0$) the energy distribution
returns to approximately that of the non-interacting gas.  Frequencies persist
into the classically excluded region as a result of the familiar evanescent
decay of energy eigenmodes into the potential wall.  The effect of the
mean-field interaction on the energy spectrum at this large radii, as can be
seen in the dark gray (red) line in Fig.~\ref{fig:power_and_auto1}(b), is much
less dramatic (e.g., only frequencies $\omega>\mu_\mathrm{c}/\hbar$ are
present), and the population of frequencies is only significant up to a small
increment above the cutoff energy, which we interpret as the mean-field shift of
the highest energy single-particle modes.  We measure this shift to be
$\approx2.5\%$, indicating that the cutoff we have chosen is sufficiently high
to contain all modes significantly modified by the condensate's presence (see
Sec.~\ref{sec:Formalism}).

\begin{figure}
	\includegraphics[width=0.45\textwidth]{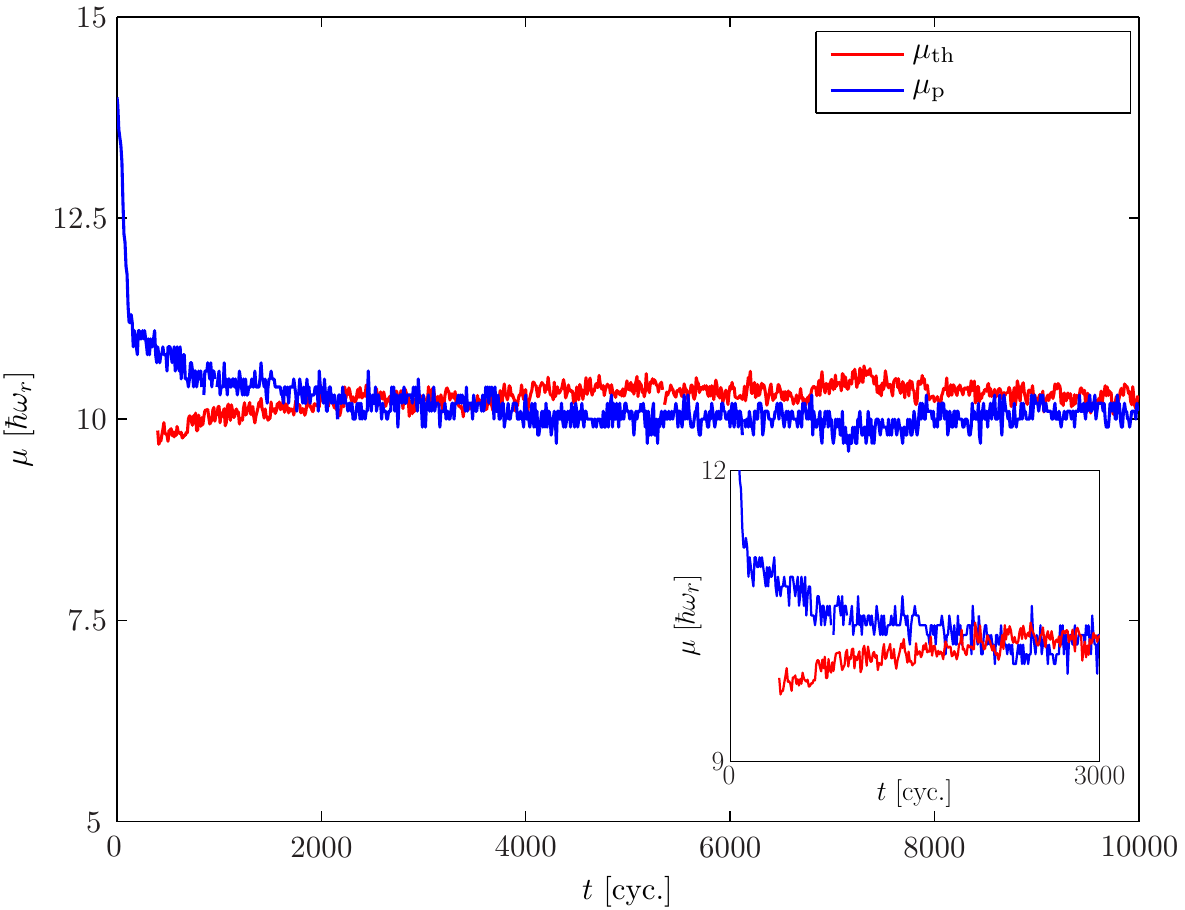}
	\caption{\label{fig:mu_comparison} (Color online) Effective chemical
	potential of the condensate and thermal cloud.  The data displayed is
	obtained from 10 trap cycle long samples spaced at 100 trap cycle
	intervals.  Inset: Data with samples spaced at 10 trap cycle intervals
	over the initial decay (rise) of $\mu_\mathrm{p}$ ($\mu_\mathrm{th}$).}
\end{figure}

In order to follow the relaxation of the condensate, we measure the frequency
$\omega_\mathrm{p}$ of maximum occupation in the power spectra, for power
spectra evaluated over various periods.  Away from equilibrium this frequency
does not necessarily correspond to the condensate eigenvalue, as many strong
collective excitations are present.  Nevertheless it provides a measure of the
energy of condensate atoms, and we interpret it as representing an effective
condensate chemical potential $\mu_\mathrm{p}=\hbar\omega_\mathrm{p}$.  In
Fig.~\ref{fig:mu_comparison} we plot the condensate chemical potential obtained
in this way and also the chemical potential of the thermal cloud obtained using
our fitting procedure (Section \ref{subsec:fitting}) for comparison.  We see
that the condensate chemical potential reduces with time, as interactions with
thermally occupied modes damp its excitations.  By $t\approx 4000$ cyc., the two
chemical potentials appear to have reached diffusive equilibrium, to the
accuracy of our fitting procedure for $\mu$ and $T$.

\subsubsection{Temporal coherence and condensate identification} The power
spectrum discussed above shows clearly that a condensate is present within the
classical field, presenting as its signature the presence of a single highly
occupied frequency, broadened by its interactions with the other modes in the
system.  The presence of this narrow frequency spike is indicative of the
temporal coherence of this condensate mode, and we will use the local
autocorrelation function \begin{equation}
	G(\mathbf{x},\tau;t_0) =
	[\mathfrak{F}_{t_0}^T]^{-1}\{H(\mathbf{x},\omega;t_0)\},
\end{equation} of the field at point $\mathbf{x}$, at lag $\tau$ relative to the
time $t_0$ to quantify the local temporal coherence of the field.  To do so we
employ the Wiener-Khinchin theorem \cite{Press92} to evaluate the local
autocorrelation function.

In practice we take the discrete Fourier transform of
$\tilde{H}(r_k,\omega;t_0)$ to obtain the azimuthally averaged autocorrelation
$\tilde{G}(r_k,\tau;t_0)$, and normalise it so that $|\tilde{G}(r_i,\tau=0,t_0)|
= 1$ for all radii $r_i$.  We will use the modulus $|\cdots|$ to characterise
the temporal coherence of the classical field, as a function of position, and
take the timescale for its decay to represent the timescale over which the field
remains temporally coherent at radius $r_k$, near time $t_0$.  Illustrative
results are shown in Figs.~\ref{fig:power_and_auto1}(c) and (d).

At the centre of the trap (light gray/cyan line in
Fig.~\ref{fig:power_and_auto1}(d)), the peak of the autocorrelation function is
broad and decays monotonically with increasing $|\tau|$.  At larger radii
($r\approx 2r_0-5r_0$), this peak possesses a nontrivial structure (black/blue
line), with $|\tilde{G}|$ vanishing and then increasing again as $\tau$ varies.
Analyzing the trajectories of vortices during the sampling period we conclude
that this structure is due to the passage of vortices through these radii.  We
identify the width of the central peak (Fig.~\ref{fig:power_and_auto1}(d)) at a
radius $r_i$ as (twice) the correlation time of the classical field at this
radius.  At the trap centre, the correlation time $t_\mathrm{c} \approx 2.2$
cyc., indicating that despite the presence of highly energetic thermal
excitations and mode mixing in the central region, the temporal coherence of the
condensate mode is significant.  We note also that a small ridge appears in
$|\tilde{G}(\tau)|$ for $\tau$ close to zero, and interpret this as representing
the correlations of the thermal component of the field (see also
\cite{Bezett08}).

The width of the broad envelope (Fig.~\ref{fig:power_and_auto1}(d)) decreases
with increasing radius around $r=8r_0$, where the strong peak in the power
spectrum also reduces.  Past this point the peak becomes much narrower (dark
grey/red line in Fig.~\ref{fig:power_and_auto1}(d)); the short correlation time
of the outer cloud reflecting the broad range of energies characteristic of the
thermally occupied excitations.  Beyond the classical turning point, the
evanescent nature of the basis modes yields a spurious amount of temporal
coherence and this manifests as an increasing correlation time past this point
(a similar phenomenon was observed for spatial correlations in \cite{Bezett08}).

\begin{figure}
	\includegraphics[width=0.45\textwidth]{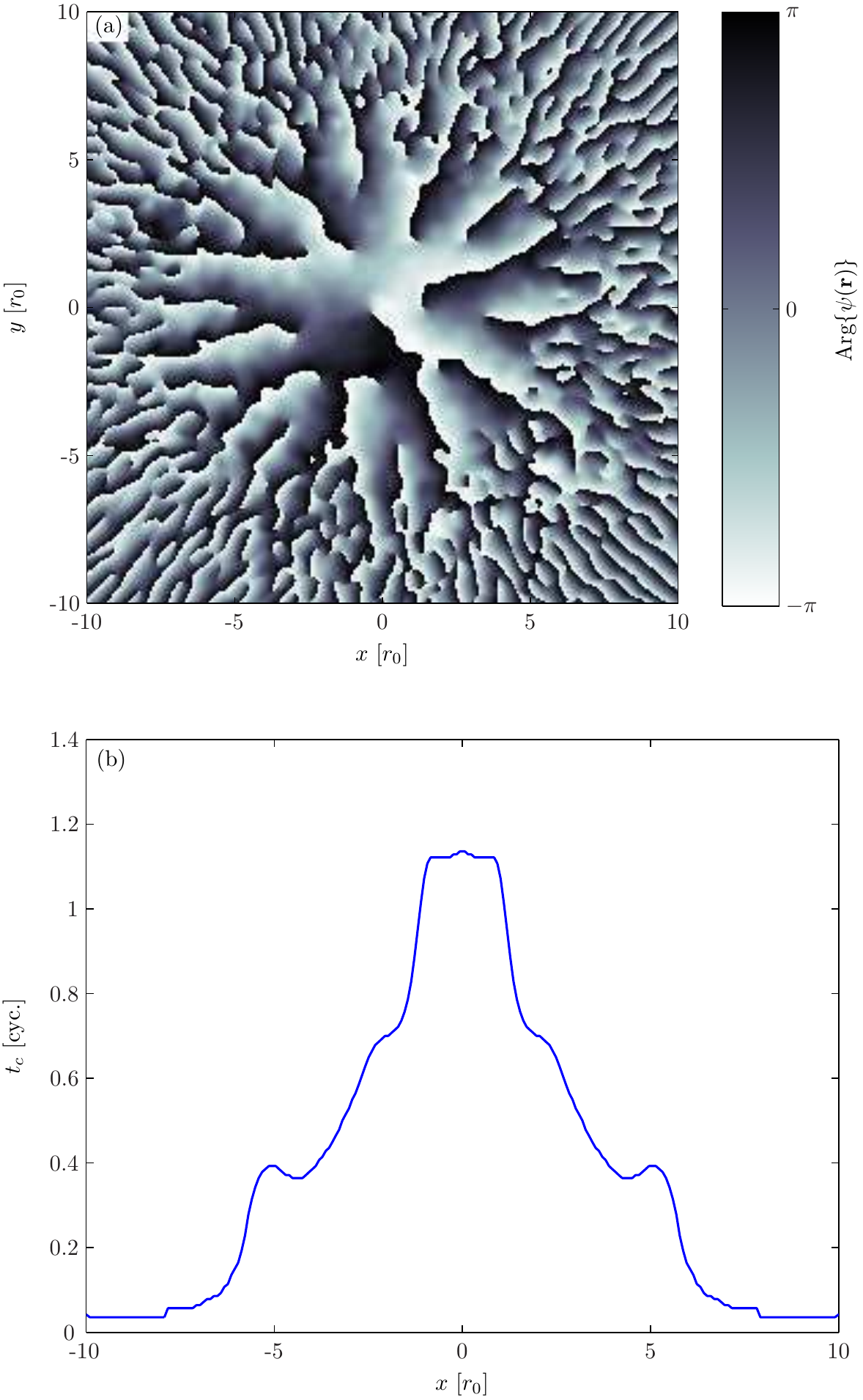}
	\caption{\label{fig:phase_and_correl_time} (Color online) Plots of (a)
	the condensate band phase and (b) the corresponding correlation time as
	a function of radius, near time $t=100$ cyc.  }
\end{figure}

While the interpretation of the correlation time we calculate in this manner is
complicated by the non-monotonic decay of $|\tilde{G}(\tau)|$ with increasing
$|\tau|$, it nevertheless allows us to distinguish \emph{turbulent} behaviour
from \emph{thermal} behaviour.  For example, in
Fig.~\ref{fig:phase_and_correl_time}(a) and (b) we plot the phase profile of
condensate band at time $t=100$ cyc, and the corresponding correlation time
respectively.  The correlation time maintains a near-constant value of
$t_\mathrm{c}\approx 1.1$ cyc.  as $r$ increases until $r\approx r_0$, where the
first vortex (phase singularity in Fig.~\ref{fig:phase_and_correl_time}(a))
occurs.  At this point the correlation time begins to decay steeply with
increasing $r$.  Throughout the region $r<r_\mathrm{tp}$ the phase structure is
complicated, however we see that the correlation time of the central region is
at least of order $t_\mathrm{c} \sim 0.5$ trap cycles, indicating that despite
the turbulent nature of the field here, interactions serve to maintain its
coherence, and we identify the behaviour here as superfluid turbulence
\cite{Barenghi01}, in contrast to the outer region ($r\gtrsim6r_0$), which has a
very short correlation time and thus appears to be truly thermal material.  This
is in contrast to the interpretation presented by the authors of
\cite{Parker05a,Parker06b}, who considered the turbulence developed during the
stirring process to be purely superfluidic (i.e., zero temperature) in nature.

\subsection{Vortex motion}\label{subsec:motional_damping} As noted in
Sec.~\ref{sec:Results}, the motion of vortices (as viewed in the rotating frame)
is initially very rapid, and slows as time goes on.  In order to quantify this
motion and its slowing with time, we track vortex trajectories, monitoring the
coordinates of all vortices within a circular region of radius equal to the
Thomas-Fermi radius of the initial lab-frame condensate.  Due to the rotational
dilation of the vortex-containing condensate, this region always lies within the
central bulk of the condensate, as defined by the temporal analysis described
above.  In general, vortices enter and leave this counting region as time
progresses.  We are interested in the motion of vortices which persist in the
counting region and can therefore be considered to exist in the condensate,
rather than those which occur in the violently evolving condensate periphery.
We therefore discard the trajectories of all vortices which do not remain in the
region for at least half the counting period of $T_\mathrm{v} = 10$ trap cycles
about time $t_i$.

\begin{figure}
	\includegraphics[width=0.45\textwidth]{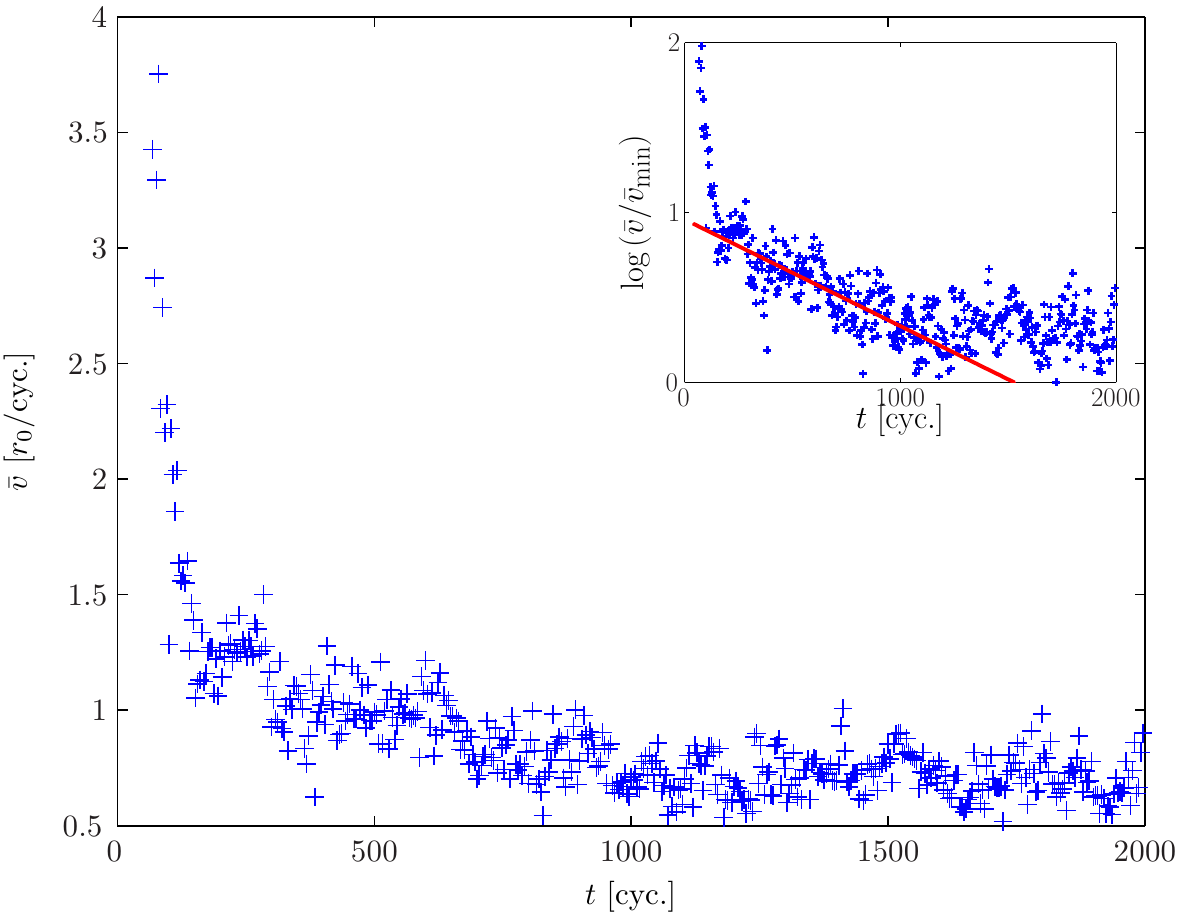}
	\caption{\label{fig:vortex_damping1} (Color online) Mean vortex velocity
	measured from the classical field trajectory, revealing the rapid
	slowing of vortices during the initial stage of nucleation
	($t\approx0-120$ cyc.), and more gradual damping of the remaining
	collective excitations ($t\approx100-1000$ cyc.).  Inset: Logarithm of
	the mean vortex velocity (scaled by the minimum velocity measured
	$\bar{v}_\mathrm{min}$), and linear least-squares fit to the data over
	the damping period.}
\end{figure}

The motion of the remaining vortices is erratic on a length scale of order of
the healing length, which is the manifestation of thermal core-filling in the
classical field model.  As this motion occurs on a short timescale, we interpret
it as the signature of high-energy excitations of the vortices.  Our interest
however is in the low energy component of vortex motion, which undergoes damping
as the field relaxes to equilibrium.  We therefore apply a Gaussian frequency
filter centred on $\omega=0$ to the trajectory components $\{x(t),y(t)\}$ to
remove their highest \emph{temporal} frequency components.  We choose the width
of this filter to be $\sigma_\omega \sim 0.6\omega_r$, and so the frequency
components preserved in the trajectories correspond to low energy excitations
($\omega \sim \omega_r$).  From the $N_t$ coordinate pairs $(x_k,y_k)$ of the
$j^\mathrm{th}$ filtered vortex trajectory we extract the mean speed along the
trajectory arc, \begin{equation}\label{eq:arc_speed}
	v_j(t_i) = \frac{1}{T_\mathrm{v}}\int ds \approx \frac{1}{T_\mathrm{v}}
	\sum_{i=1}^{N_t - 1} \sqrt{(x_i-x_{i-1})^2 + (y_i - y_{i-1})^2}.
\end{equation} We then average this quantity over the vortices included in the
count, to find the mean speed per vortex $\overline{v}(t_i) =
\sum_{j=1}^{N_\mathrm{v}} v_j(t_i)/N_\mathrm{v}$.  In
Fig.~\ref{fig:vortex_damping1} we plot this quantity as a function of time.  We
note firstly that the mean vortex speed decays dramatically during the initial
measuring period $t\approx70-150$ cyc., as vortices enter the condensate with
high velocities (with a substantial component opposing the trap rotation, as
observed in the rotating frame), and their motion is damped heavily by their
interaction with one another and with the thermal component.  Subsequent to
this, a secondary, slower damping is observed over the period $t\approx
150-1100$ cyc., after which the vortex motion seems to have reached a lower
limit set by thermal excitations.  During this period the condensate contains
nearly the number of vortices it contains in equilibrium, and we interpret the
damping during this period as the damping of low energy lattice excitations due
to their interaction with thermally occupied, high energy modes.  Taking the log
of the speed data (Fig.~\ref{fig:vortex_damping1} inset), we see that the decay
during this period is approximately exponential, and performing a linear fit to
the data over this period we extract a damping rate $\gamma \approx
6.3\times10^{-4}$ $(\mathrm{cyc.})^{-1}$.

We note that the final distribution of vortices is disordered, with the vortices
failing to become localised in a regular lattice.  This result is consistent
both with the considerations of Sec.~\ref{subsec:Thermo_params} and with pure
GPE studies of stirring in 2D performed by Feder \emph{et al.} \cite{Feder01a},
and Lundh \emph{et al.} \cite{Lundh03}.  The results presented here (and those
of \cite{Lobo04}) suggest that the turbulent states observed in
\cite{Feder01a,Lundh03} are in fact finite temperature ones beyond the validity
of the GP model employed in those calculations.  By contrast, the calculations
of Parker \emph{et al.} \cite{Parker06b,Parker05a} produced fully crystallized
vortex lattices in 2D, indicating that additional damping mechanisms were
available to the system in their simulations.  It is possible that these are
attributable to the spatial differencing technique \cite{Parker_PhD,note13} used
in those calculations, which is known to be less accurate \cite{note14} than the
methods used in the present work, and Refs.  \cite{Feder01a,Lundh03}.

%%%%%%%%%%%%%%%%%%%%%%%%%%%%%%%%%%%%%%%%%%%%%%%%%%%%%%%%%%%%%%%%%%
\section{Dependence on simulation parameters}\label{sec:parameters} We compare
now the behaviour of stirred condensates for different condensate sizes, i.e.,
for different values of the initial chemical potential $\mu_\mathrm{i}$.  The
field density and thus the nonlinear effects of mean-field interaction increase
as the chemical potential is increased, and larger condensates in rotational
equilibrium will also contain larger numbers of vortices.  In
Sec.~\ref{subsec:vary_mu} we assess the effect of varying initial chemical
potential on our classical field solutions.

It is also important in performing classical field calculations such as those
presented here to consider the effect of the cutoff height on the results of
simulations.  In Appendix~\ref{app:AppendixB} we compare results for simulations
with fixed initial chemical potential but varying cutoff heights, and quantify
the effect of this purely technical parameter on the physical predictions of our
model.

\subsection{Chemical potential}\label{subsec:vary_mu} The initial chemical
potential impacts upon the system evolution both qualitatively, affecting the
density profile of the condensate mode and thus the nature of the dynamical
instability leading to vortex nucleation, and quantitatively, determining the
equilibrium parameters of the stationary state of the classical field.  The
characteristics of the vortex array and dynamics of the vortices are also
strongly dependent on the chemical potential.  We consider these three aspects
in turn.  \subsubsection{Effect on dynamical instability} 
\begin{figure}
	\includegraphics[width=0.45\textwidth]{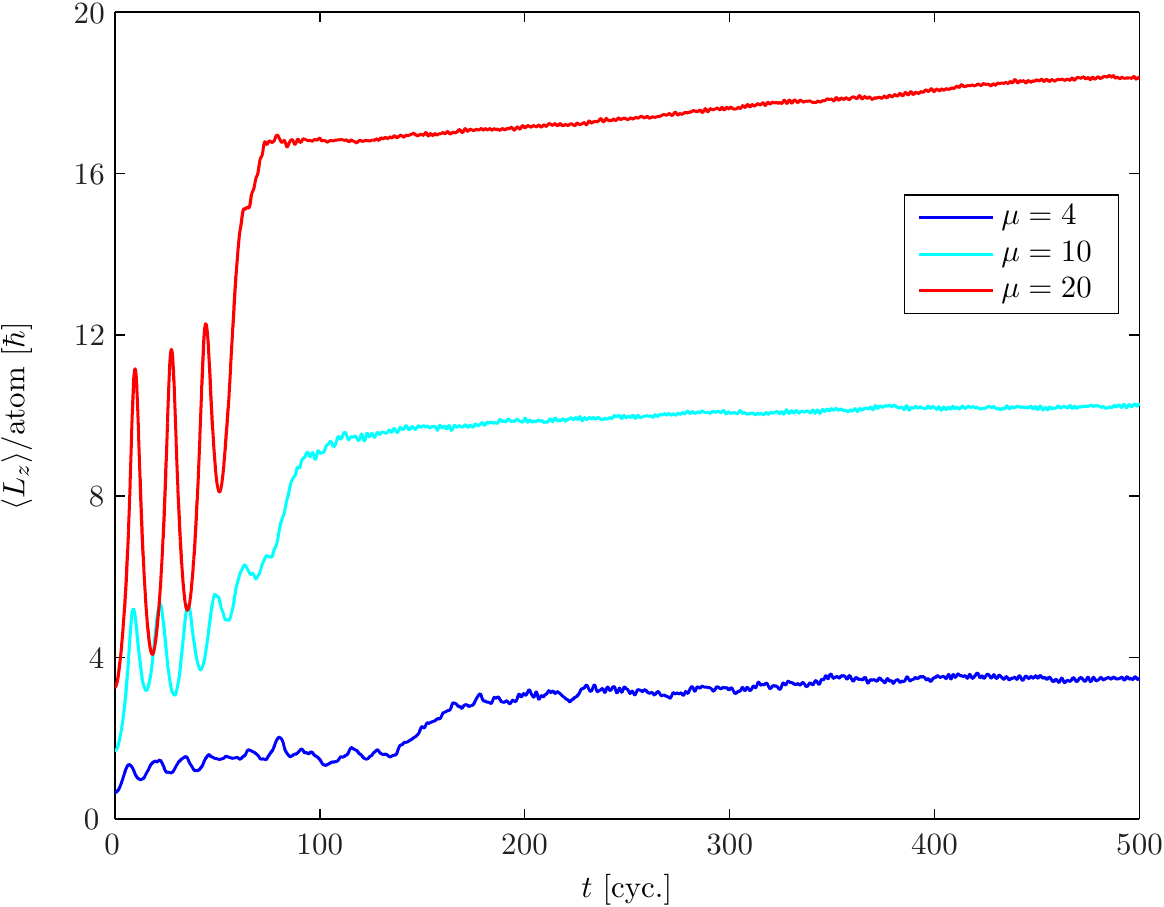}
		\caption{\label{fig:Lz_different_mu} (Color online) Evolution of
		the field angular momentum for various initial chemical
		potentials.}
	\end{figure}

The chemical potential
determines the strength of the mean-field repulsion influencing the condensate's
shape.  As the chemical potential becomes large, the condensate mode approaches
that described by the Thomas-Fermi approximation, whereas for smaller chemical
potentials, the kinetic energy of the condensate becomes important, and the
condensate boundary broadens.  Consequently the visual distinction between the
condensate mode and its incoherent excitations becomes less clear as the
chemical potential is reduced.  This affects the early evolution of the system
and the onset of the dynamical instability in a pronounced manner.  For the
largest chemical potentials considered ($\mu \approx 17-20\hbar\omega_r$), the
collective (quadrupole) excitations and their breakdown are clearly visible
against a background of incoherent noise resulting from the initial vacuum
occupation.  For smaller chemical potentials, while quadrupole oscillations are
still visible and the outer cloud becomes visibly more dense as time proceeds,
indicating that material is indeed lost from the condensate, any ejection of
material from the condensate is obscured by the blurred condensate boundary.
Persistent `ghost' vortices can be seen forming in the incoherent region about
the condensate periphery very early in the evolution ($t\approx 20$ cyc.  for
$\mu_\mathrm{i}=4\hbar\omega_r$).  In Fig.~\ref{fig:Lz_different_mu} we plot the
evolution of the field angular momentum per atom for three different initial
chemical potentials.  The most obvious feature is that larger condensates
support more angular momentum per atom, due to increased mean-field effects.
The evolution of the field angular momentum for the smallest condensate
considered ($\mu_\mathrm{i}=4\hbar\omega_r$) reveals the onset of irreversible
coupling of angular momentum into the field at $t\approx 130$ cyc.  Inspection
of the field density evolution shows that vortices begin to be nucleated into
the central region of the field at about this time.  We therefore conclude that
despite the absence of visible break-up of the condensate, the unstable
quadrupole oscillations still serve to produce the thermal component required to
allow vortices to enter the central condensate region \cite{note10}.

\subsubsection{Thermodynamic parameters} 
\begin{figure}
	\includegraphics[width=0.45\textwidth]{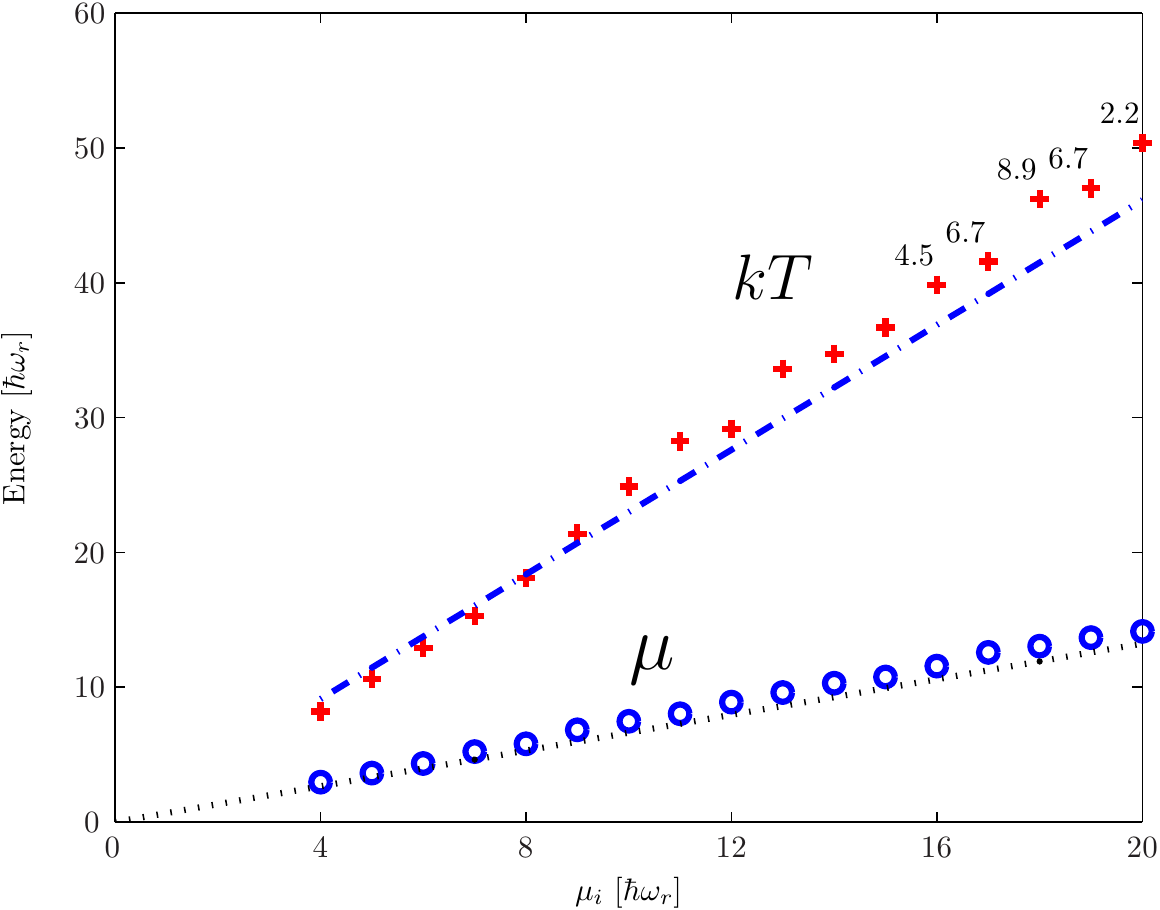}
	\caption{\label{fig:thermo_variation_with_mu} (Color online) Equilibrium
	temperatures and chemical potentials reached as a function of initial
	chemical potential, as determined by the fitting procedure.  Dotted and
	dash-dot lines indicate the predictions of Eq.~(\ref{eq:reduced_mu}) and
	Eq.~(\ref{eq:temperature_prediction}) respectively.  Numbers indicate
	times at which the measurements were made (thousands of trap cycles) in
	cases where numerical results were constrained by time.  Points without
	numbers were measured at $10^4$ cyc.}
\end{figure}

As described in
Section~\ref{subsec:Thermo_params}, a simple analysis based on energetic
considerations allows us to predict the equilibrium temperatures and chemical
potentials of our classical field simulations, and the dependence of these
parameters on the initial chemical potential and the condensate band
multiplicity of our simulations.  In Fig.~\ref{fig:thermo_variation_with_mu} we
plot these analytic predictions alongside the values obtained using the fitting
procedure of Section~\ref{subsec:Thermo_params}.  The dotted line shows the
estimated chemical potential of the (idealised) norm conserving, rotating frame
condensate mode, $\mu_\Omega = \mu_\mathrm{i}\sqrt{1-\Omega^2/\omega_r^2}$,
while the dash-dot line indicates the estimated equilibrium temperature given by
Eq.~(\ref{eq:temperature_prediction}).  We note that despite its approximate
(TF-limit) and asymptotic (zero thermal fraction) nature, the analytical
estimate for $\mu$ is a reasonable prediction of the measured values.  The
measured value of $\mu$ is slightly higher than the TF prediction, and this
discrepancy appears to increase with increasing $\mu_\mathrm{i}$, while the
measured temperature appears consistently smaller than the analytical prediction
at small $\mu_\mathrm{i}$, and exceeds it as the initial chemical potential is
increased.  However, the agreement seems very reasonable given the simplicity of
the arguments presented in Section~\ref{subsec:Thermo_params}, which neglect
both the kinetic energy of vortices (TF approximation), the depletion of the
condensate population and all other effects beyond a linear (Bogoliubov)
description.  We therefore conclude that our simple analysis captures the
essential physics determining the thermodynamic parameters of the equilibrium.
That is to say, the temperature is determined by the necessity of the system to
redistribute the excess energy of the initial state (as viewed in the rotating
frame), and that no additional heating of the system is required for the system
to migrate to a state containing vortices.

\subsubsection{Vortex array structure} Perhaps the most striking difference
between vortex arrays in simulations of differing condensate populations is the
number of vortices present.  As noted in Sec.~\ref{subsec:Rotational_params},
the density of vortices is largely determined by the rotation rate of the
condensate and is given to leading order by Eq.~(\ref{eq:Feynman_relation}), and
(within the TF approximation) the total number of vortices in an equilibrium
condensate is approximately proportional to the chemical potential $\mu$.

The stability of a vortex array in equilibrium is determined by the competition
between hydrodynamic and energetic considerations \cite{Donnelly91} which favour
the crystallization of a rigid lattice, and the perturbing effect of thermal
fluctuations in the atomic field.  Close to zero temperature these fluctuations
are interpreted as the thermal occupation of Bogoliubov excitations of the
underlying vortex lattice state.  As the temperature of a vortex lattice state
is increased it eventually undergoes a transition to a disordered state
\cite{Bradley08} as the long-range order of the lattice is degraded by thermal
fluctuations.

After a long time period, $t \sim 10^4$ trap cycles, all our simulated fields
appear to describe states on the disordered side of this transition.  All our
solutions have vortices migrating within the condensate bulk, and leaving and
entering through the condensate periphery.  In the the smallest condensates,
with the smallest vortex counts (i.e.  $\mu = (4,5,6)\hbar\omega_r$ with counts
$N_\mathrm{v} \approx (4,5,6)$), the vortex array is small enough that
\emph{quasi-regular} configurations of vortices may occur, in which the vortex
positions appear to fluctuate about approximate equilibrium positions on a
triangular (or square) lattice.  Such configurations may persist for as long as
$\sim 10$ trap cycles, but ultimately breakdown and give way to new
configurations as vortices cycle in and out of the central, high density region
of the field.  We interpret this cycling behaviour as the classical field
`sampling' different configurations during its ergodic evolution \cite{note11}.

\begin{figure}
	\includegraphics[width=0.45\textwidth]{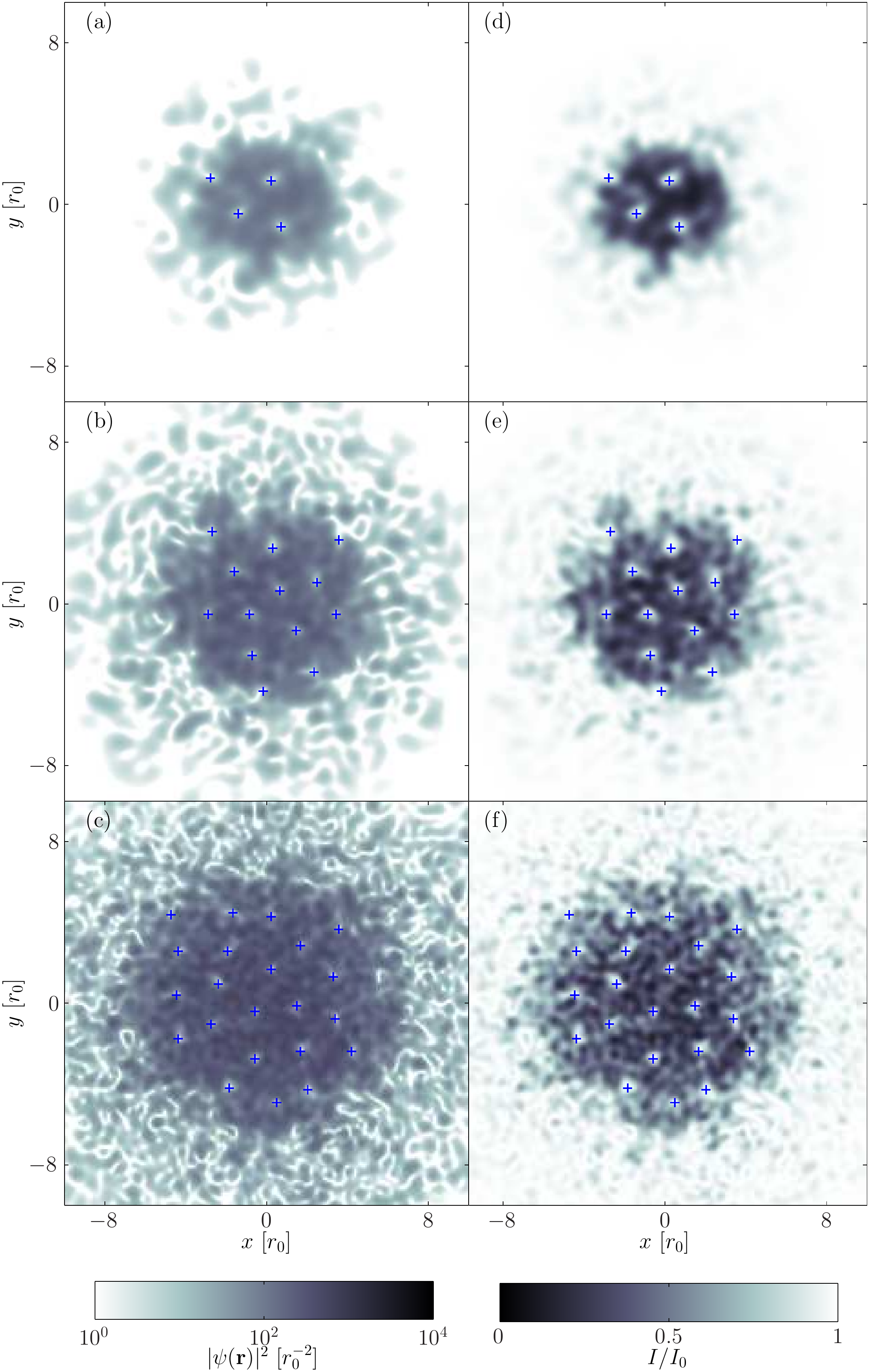}
	\caption{\label{fig:densities_vs_mu} (Color online) (a-c) Coordinate
	space densities of the classical fields with initial chemical potentials
	$\mu_\mathrm{i}=5,10,18\hbar\omega_r$ at times $t=9900,9900,8800$ cyc.
	respectively.  (d-f) Simulated absorption images generated from the same
	densities.}
\end{figure}

Larger vortex arrays appear to be prohibited from forming regular structures
over scales larger than nearest-neighbour inter-vortex separations (the scale of
the apparent order in the smaller condensates).  As noted in
Sec.~\ref{subsec:Thermo_params}, the equilibrium temperature attained by our
simulated atomic fields increases linearly with the initial chemical potential,
and so it seems unlikely that this behaviour would abate when the condensate
size is increased further.  Fig.~\ref{fig:densities_vs_mu}(a-c) we plot the
coordinate space densities of vortex arrays of differing sizes, corresponding to
differing initial chemical potentials.  Because of their small size compared to
the extent of the condensate, vortex cores are difficult to experimentally image
\emph{in situ}, and so their presence is typically detected after free expansion
of the atom cloud \cite{Lundh98} by absorptive imaging techniques.  For
condensates in the TF regime the expansion is well described by a simple scaling
of the position-space density \cite{Kagan96,Castin96} (though the vortex core
radius grows in somewhat greater proportion \cite{Lundh98}).  The Beer-Lambert
law \cite{Robinson96} for the absorption of an optical probe, and the assumption
of a ground-state (Gaussian) profile in $z$ yields for the transmitted intensity
of the probe \begin{equation}\label{eq:transmitted_intensity}
	I(x,y) = I_0e^{-\gamma|\psi(x,y)|^2},
\end{equation} with $I_0$ the input probe intensity and $\gamma$ a constant
which depends, in general, on the absorptance of the atomic medium.  In
Figs.~\ref{fig:densities_vs_mu}(d-f) we plot simulated intensity profiles
obtained using Eq.~(\ref{eq:transmitted_intensity}), where we have in each case
simply set $\gamma = 2/\{|\psi|^2\}_\mathrm{max}$, with
$\{|\psi|^2\}_\mathrm{max}$ the peak value of the density occurring in the
distribution, so as to best represent the significant features.  This process
accentuates both the disparity in density between the condensate and the outer
thermal cloud, and the density fluctuations in the central condensate region, in
contrast to the logarithmic plots (Figs.~\ref{fig:densities_vs_mu}(a-c)) which
suppress them.  The images show the thermally-distorted vortex array structure
that might be measured in an experiment.  The fact that the formation of rigid
vortex lattices here is inhibited by thermal fluctuations, while ordered
lattices have been formed experimentally at comparatively high temperatures
\cite{Abo-Shaeer02, Madison00}, suggests that this inhibition of lattice order
may be a manifestation of the well-known susceptibility of two-dimensional
system to long-wavelength fluctuations \cite{Posazhennikova06}.  Indeed (to our
knowledge), no investigations of the effect of rotation on the BEC and BKT
phases of Bose gases in quasi-2D trapping geometries have been performed to date
\cite{note12}.

\subsubsection{Damping of vortex motion} 
\begin{figure}
	\includegraphics[width=0.45\textwidth]{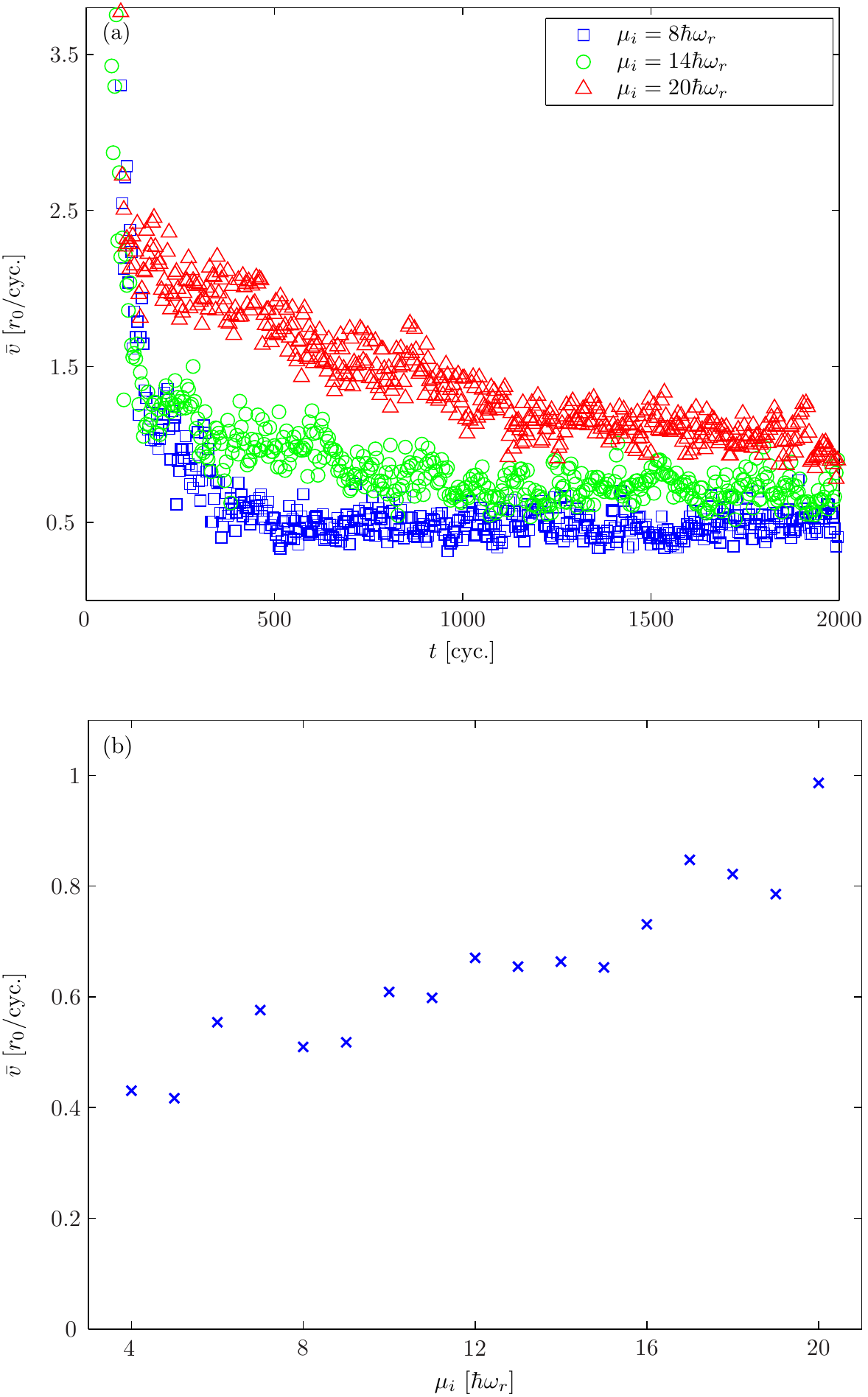}
	\caption{\label{fig:motional_mu} (Color online) (a) Vortex motional
	damping for 3 representative choices of initial chemical potential
	$\mu_\mathrm{i}$.  (b) Mean vortex motion (averaged over the period
	$t=1900-2000$ cyc.  as a function of initial chemical potential
	$\mu_\mathrm{i}$.  }
\end{figure}

While the vortex motion as measured by
the mean velocity defined in Sec.~\ref{subsec:motional_damping} exhibits the
same overall damping behaviour in all our simulations, the damping appears to be
generally non-exponential and quite sensitive to the initial noise conditions.
We have observed additional features (e.g.  transient and oscillatory behaviours
superposed with the damping) peculiar to each trajectory.  It is therefore
difficult to unambiguously quantify the rates of motional damping in order to
compare the dependence on initial chemical potential, without generating
ensemble data for each parameter set, which would be a very heavy task
numerically.  We therefore plot in Fig.~\ref{fig:motional_mu}(a) only the
behaviour of three particular trajectories, which indicate at a qualitative
level that the vortex motion in the smaller condensates damps to equilibrium
more quickly.  We note however that the level of vortex motion that each
trajectory damps to shows a clear dependence on the initial chemical potential.
In Fig.~\ref{fig:motional_mu}(b) we plot the mean velocity of vortices over the
period $t=1900-2000$ cyc., at which point the motion in each simulation appears
to have essentially relaxed to its equilibrium level.  The mean velocity so
measured clearly increases with chemical potential.  After comparing the results
for varying cutoff at fixed initial chemical potential (not shown) we conclude
that this is likely due to the increase in equilibrium temperature with
increasing initial condensate size.

%%%%%%%%%%%%%%%%%%%%%%%%%%%%%%%%%%%%%%%%%%%%%%%%%%%%%%%%%%%%%%%%%%
%%%%%%%%%%%%%%%%%%%%%%%%%%%%%%%%%%%%%%%%%%%%%%%%%%%%%%%%%%%%%%%%%%
\section{Conclusions} We have carried out the first strictly Hamiltonian
simulations of vortex nucleation in stirred Bose-Einstein condensates.  Our
approach is free from grid method artifacts such as aliasing and spurious
damping at high momenta, enabling a controlled study of the thermalization and
vortex nucleation within an explicitly Hamiltonian classical field theory.

\emph{Vacuum symmetry breaking.}---The importance of symmetry breaking has been
highlighted in previous works~\cite{Penckwitt02,Parker05a,Parker06b}.  Sampling
of initial vacuum noise provides an irreducible mechanism for breaking the
twofold rotational symmetry of the classical field solution.

\emph{Thermalization.}---Resonant excitation of the quadrupole instability
evolves the system from a zero temperature initial state through a dynamical
thermalization phase in which a rotating thermal cloud forms.  Subsequent vortex
nucleation in the remaining condensate shows that the cloud provides the
requisite dissipation to evolve the condensate subsystem toward a new
quasi-equilibrium state in the rotating frame.  This picture is consistent both
with energetic constraints of a time independent rotating frame Hamiltonian and
with previous treatments based on introducing dissipation into the GP
equation~\cite{Tsubota02a,Penckwitt02}.

\emph{Identifying the thermal cloud.}--- We have quantified the heating caused
by dynamical instability through a range of measures.  Spectral analysis of
ergodic classical field evolution has been used to identify the condensate
chemical potential, and thermal cloud properties were extracted using a
semiclassical fit at high energies.  While the chemical potentials of the
condensate and emerging thermal cloud are slower to equilibrate, the moments of
inertia, temperature, and angular momentum of the classical field are rapidly
transformed by the instability to those of a rotating, heated Bose gas.

\emph{Frustrated crystallization.}---Vortices are seen to nucleate and enter the
bulk of the condensate but a regular Abrikosov lattice does not form.  Instead
the vortices distribute throughout the condensate in spatially disordered vortex
liquid state, consistent with substantial thermal excitation of an underlying
regular vortex lattice state.  In this respect our results are consistent with
previous 2D GPE treatments of similar stirring configurations.

We conclude that vortex nucleation by stirring is a finite temperature effect,
which arises from dynamical thermalization when initiated from a zero
temperature BEC.  While we observe vortex \emph{nucleation} in 2D, the
asymptotic absence of lattice rigidity suggests that the temperature attained in
accommodating the excess (rotating frame) energy of the initial state is such
that vortex lattice \emph{crystallization} is inhibited in this minimal,
conserving stirring scenario in 2D.

%%%%%%%%%%%%%%%%%%%%%%%%%%%%%%%%%%%%%%%%%%%%%%%%%%%%%%%%%%%%%%%%%%
%%%%%%%%%%%%%%%%%%%%%%%%%%%%%%%%%%%%%%%%%%%%%%%%%%%%%%%%%%%%%%%%%%
\begin{acknowledgments} We wish to acknowledge useful discussions with M.  J.
Davis, C.  Lobo, and M.  Gajda.

This work was supported by the New Zealand Foundation for Research, Science and
Technology under Contract Nos.  NERF- UOOX0703, Quantum Technologies, Marsden
Contract No.  UOO509 and The Australian Research Council Centre of Excellence
for Quantum-Atom Optics.  \end{acknowledgments}
%%%%%%%%%%%%%%%%%%%%%%%%%%%%%%%%%%%%%%%%%%%%%%%%%%%%%%%%%%%%%%%%%% \appendix
\section{Calculation of the trap anisotropy}\label{app:AppendixA} In this
appendix we outline our numerical method for calculating the dimensionless
perturbing potential $V_\epsilon(\bar{r}) = -\frac{\epsilon}{2} \bar{r}^2
\cos(2\theta)$.  To calculate this term we follow \cite{Bradley08} in exploiting
the properties of the Laguerre-Gaussian basis modes to evaluate this
perturbation \emph{exactly} using a Gauss-Laguerre quadrature rule.  Changing
variables Eq.~(\ref{eq:anisotropy}) becomes
\begin{eqnarray}\label{eq:H_int_vars_changed}
	H_{nl} (\psi)&=& -\frac{\epsilon}{4\pi} \int_0^{2\pi} d\theta
	e^{-il\theta} \cos(2\theta) \\ \nonumber &&\times \int_0^\infty dx
	x\Phi_{nl}(x)\psi(\sqrt{x},\theta)\pi,
\end{eqnarray} where $\Phi_{nl}(x) = Y_{nl}(\sqrt{x},\theta) e^{-il\theta}$.
The integral to evaluate is therefore of form
\begin{equation}\label{eq:integral_form}
	H_{nl}(\psi) = \frac{1}{4\pi} \int_0^{2\pi} d\theta e^{-il\theta}
	\cos(2\theta) \int_0^\infty dx e^{-x}Q(x,\theta),
\end{equation} where $Q(x,\theta)$ is a polynomial in $x$ and $e^{i\theta}$ of
order determined by the cutoff.  The order of $e^{i\theta}$ in
$\psi(\mathbf{x})$ is constrained by Eq.~(\ref{eq:band_spectrum}) to be
$-\overline{l}_- \equiv -l_-(0) \leq l \leq l_+(0) \equiv \overline{l}_+$, where
$l_\pm(n) = [(\overline{N}-2n)/(1\mp\Omega)]$, with $[...]$ denoting the floor
function.  Choosing the $\theta$ grid so as to accurately compute all integrals
$\frac{1}{2\pi}\int_0^{2\pi} d\theta e^{-iq\theta}$, where $q$ is an integer in
the range $-(\overline{l}_- + \overline{l}_+ + 2) \leq q \leq (\overline{l}_- +
\overline{l}_+ + 2)$, the integral can be carried out exactly for the condensate
band by discretizing $\theta$ as $\theta_j = j\Delta \theta = j2\pi/N_\theta$,
for $j=0,1,\cdots,N_\theta-1$, where $N_\theta = \overline{l}_- + \overline{l}_+
+ 3$, to give \begin{equation}
	\int_0^{2\pi} d\theta e^{-il\theta} \cos(2\theta) =
	\frac{2\pi}{N_\theta} \sum_{j=0}^{N_\theta-1} e^{-i2\pi j l/N_\theta}
	\cos(2\theta_j) Q(x,\theta_j).
\end{equation} The $x$-integral is computed by noting that the polynomial part
of the $x$-integrand when the angular integral is nonzero is of maximum order
$[\overline{N}/(1-\Omega)]\equiv \overline{l}_+$, which can be computed exactly
using a Gauss-Laguerre quadrature rule of order $N_x\equiv[\overline{l}_+/2]+1$.
The radial integral in Eq.~(\ref{eq:integral_form}) is \begin{equation}
	\int_0^\infty dx Q(x,\theta) e^{-x} = \sum_{k=1}^{N_x} w_k
	Q(x_k,\theta),
\end{equation} where the $x_k$ are the roots of $L_{N_x}(x)$ and the weights
$w_k$ are known constants \cite{Cohen73}.  We can thus write
Eq.~(\ref{eq:H_int_vars_changed}) as \begin{eqnarray}
	H_{nl}(\psi)&=& -\frac{\epsilon}{2} \sum_{k=1}^{N_x} \tilde{w_k}
	\Phi_{nl}(x_k) \frac{1}{N_\theta} \sum_{j=0}^{N_\theta-1} e^{-i2\pi
	jl/N_\theta} \\ \nonumber
&&\times\cos(4\pi j l / N_\theta) \psi(\sqrt{x_k},\theta_j), \end{eqnarray}
where the effective weights $\tilde{w}_k = \pi w_k x_k$.  Once the modes are
precomputed at the Gauss points \begin{equation}
	P_{kn}^{|l|} = \Phi_{n|l|} (x_k),
\end{equation} the mixed quadrature rule for calculating the perturbation due to
the trap anisotropy is: \begin{enumerate} \item Compute the radial transform
\begin{equation}
	\chi_{kl} = \sum_{n=0}^{[\overline{N}/2]} P_{kn}^{|l|} c_{nl}.
\end{equation} \item Construct the position field at the quadrature points
$\Psi_{kj}$ using the FFT, after zero padding $\chi_{kl}$ to length $N_\theta$
in the $l$ index, $\tilde{\chi}_{kl} \equiv \mathrm{pad}_l
(\chi_{kl},N_\theta)$, \begin{equation}
	\Psi_{kj} = \sum_{l=0}^{N_\theta-1} e^{i2\pi jl/N_\theta}
	\tilde{\chi}_{kl}.
\end{equation} \item Apply the perturbing potential \begin{equation}
	\Gamma_{kj} = \Psi_{kj} x_k \cos(2\theta_j).
\end{equation} \item Compute the inverse FFT \begin{equation}
	\Theta_{kl}= \frac{2}{N_\theta}\sum_{j=0}^{N_\theta-1} \Gamma_{kj}
	e^{-i2\pi jl/N_\theta}.
\end{equation} \item Calculate the Gauss-Laguerre quadrature \begin{equation}
	H_{nl} (\psi)= -\frac{\epsilon}{2} \sum_{k=1}^{N_x} \tilde{w}_k
	\Theta_{kl} P_{kl}^{|l|}.
\end{equation} \end{enumerate} The orders of the quadrature rules used here are
significantly smaller than those used for the evaluation of the nonlinear term
(see \cite{Bradley08}), and consequently the computational load increase due to
the inclusion of the trap anisotropy over that of the base method is of order
25-50\%, depending on the rotation rate and cutoff height.

\section{Cutoff dependence}\label{app:AppendixB} Of crucial importance in
classical field simulations is the effect of the basis size upon the system
behaviour.  As noted by other authors \cite{Lobo04, Sinatra02}, for simulations
in which the developed thermal fraction is small, such that a Bogoliubov-level
description of the equilibrium non-condensate distribution is appropriate, the
temperature is expected to be inversely proportional to the modes available for
thermalization.  It might be expected \cite{Lobo04} that this will influence
relaxation rates, which depend strongly on the thermal occupation of system
excitations (see e.g.  \cite{Fedichev98}).  \subsubsection{Thermodynamic
parameters} \begin{figure}
	\includegraphics[width=0.45\textwidth]{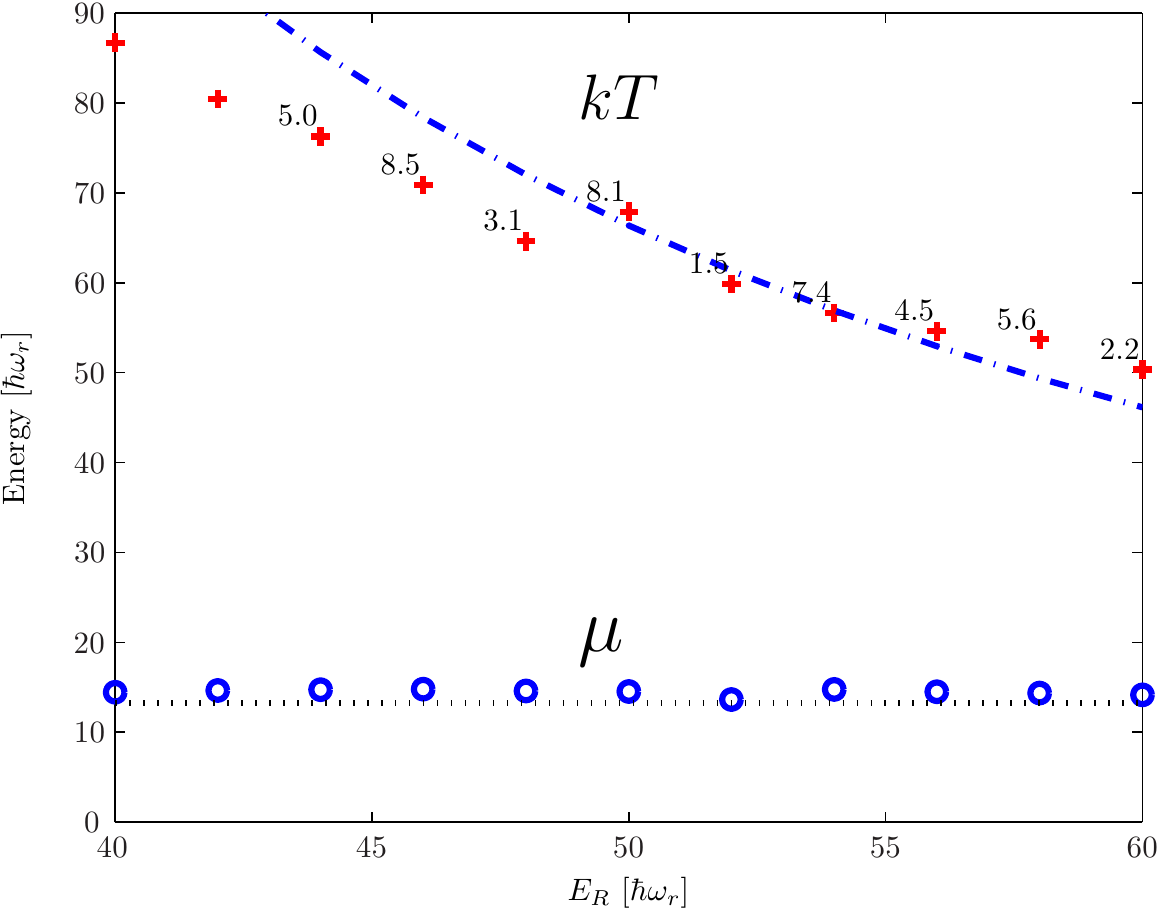}
	\caption{\label{fig:thermo_variation_with_cutoff} (Color online) Field
	temperatures and chemical potentials reached as a function of cutoff.
	Dashed and dash-dot lines indicate the predictions of
	Eq.~(\ref{eq:reduced_mu}) and Eq.~(\ref{eq:temperature_prediction})
	respectively.  Measurement times are indicated as in
	Fig.~\ref{fig:thermo_variation_with_mu}.}
\end{figure}

\begin{figure}
	\includegraphics[width=0.45\textwidth]{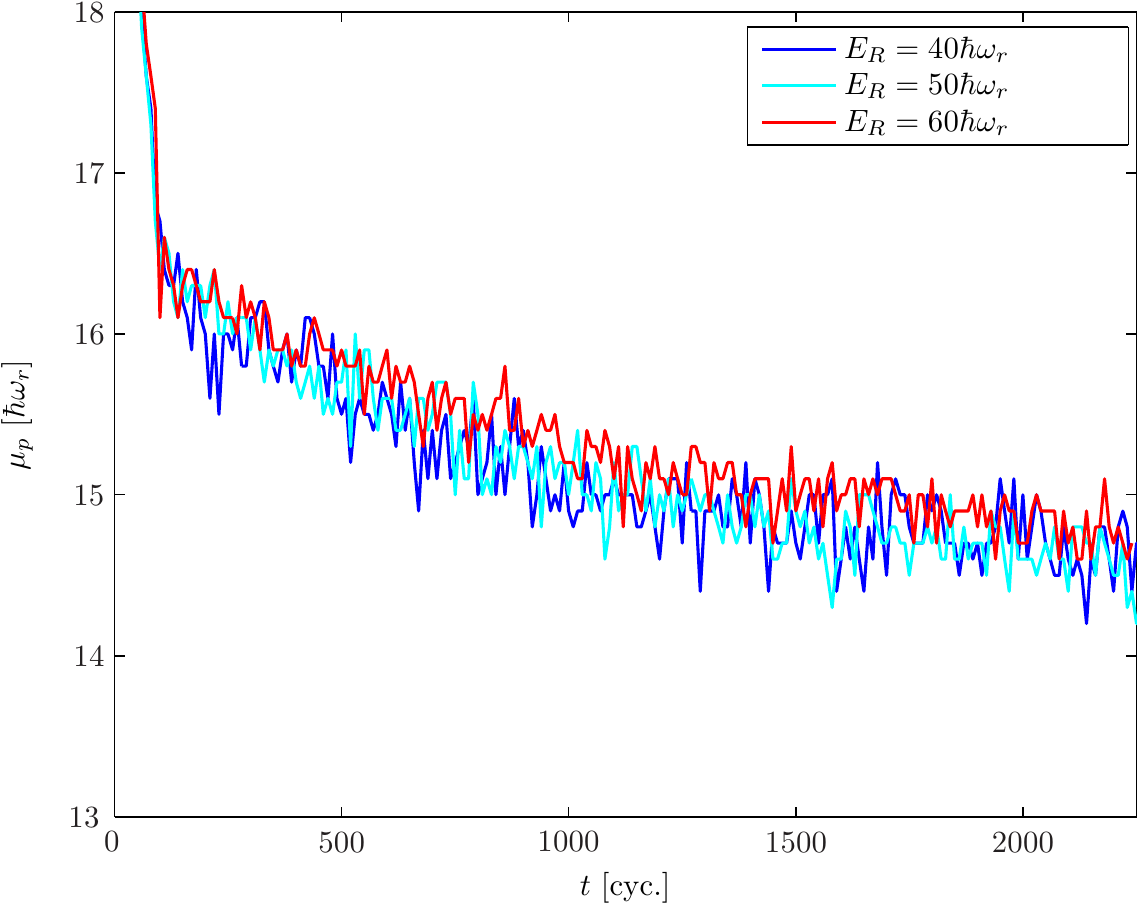}
	\caption{\label{fig:mu_damp_ER} (Color online) Decay of the effective
	condensate chemical potential (as defined in Sec.~\ref{subsec:Temporal})
	for trajectories with initial chemical potential
	$\mu_\mathrm{i}=20\hbar\omega_r$ and three different values of the
	energy cutoff $E_\mathrm{R}$.}
\end{figure}

In Fig.~\ref{fig:thermo_variation_with_cutoff} we plot the temperatures and
chemical potentials attained in simulations with $\mu_\mathrm{i}=20$ and cutoff
heights $E_\mathrm{R}$ varying over the range
$2\mu_\mathrm{i}\rightarrow3\mu_\mathrm{i}$.  We note first that while the
temperature shows a clear dependence on $E_\mathrm{R}$, the chemical potential
appears relatively insensitive to it.  The system therefore reaches a chemical
potential closest to that resulting from the idealised transformation of the
condensate to a lattice state without loss of atoms from the condensate mode.
The measured temperature decreases with increasing cutoff as expected, but
differs from the $T \propto 1/(\mathcal{M}-1)$ behaviour predicted by for a
classical field in the Bogoliubov limit.  We attribute this to the constraint
imposed by the low cutoff.  The cutoff here serves to increase the energy of
natural modes in the system, by increasing their overlap with the condensate
(c.f.  the discussion in Sec.~\ref{subsec:Temporal}), thus lowering the
equilibrium temperature of those modes.  Indeed for $E_\mathrm{R}=2\mu$, the
classical turning point $r_\mathrm{tp}\sim1.7r_\mathrm{TF}$, where
$r_\mathrm{TF} \approx 7.8r_0$ is the TF radius of the lattice state, and so the
corruption the quasiparticle mode shapes may be quite pronounced.

\subsubsection{Damping} As noted in Sec.~\ref{subsec:motional_damping}, it is
difficult to extract quantitative comparisons of relaxation rates from the
damping of vortex motion.  Here we consider the relaxation of another quantity:
the effective condensate chemical potential (power spectrum peak)
$\mu_\mathrm{p}$ introduced in Sec.~\ref{subsec:Temporal}.  In
Fig.~\ref{fig:mu_damp_ER}, we show the decay of this quantity with time for
three values of the cutoff $E_\mathrm{R}$.  We observe that the damping is
reasonably insensitive to the cutoff.  A possible explanation for this behaviour
is that as the effect of lowering the cutoff is to decrease the number of
thermally occupied modes in the system and consequently to increase the
temperature of these modes, that these two effects serve to approximately cancel
one another in determining the rates of thermal damping.

%%%%%%%%%%%%%%%%%%%%%%%%%%%%%%%%%%%%%%%%%%%%%%%%%%%%%%%%%%%%%%%%%%
\bibliographystyle{prsty}

\begin{thebibliography}{100}
\bibitem{note1} {By far the most comprehensive prescription for the addition of
noise to the
  initial states of TW-type simulations is that developed in
  \cite{Sinatra00,Sinatra01,Sinatra02}, which includes fluctuations in
  condensate occupation to second-order in the $U(1)$-symmetry preserving
  expansion of Ref.~\cite{Castin98}.  }.
\bibitem{note2} {Above cutoff corrections are included in \cite{Gardiner03}
where the dynamics
  are driven by a stationary above-cutoff thermal bath, and in \cite{Davis06}
  where only equilibrium properties of the system are considered.  Neither of
  these conditions apply in the scenario considered here.}
\bibitem{note3} {The central issue is the evaluation of the interaction term in
the GPE
  evolution.  A non-trivial modification of the method~\cite{Davis_DPhil}
  promotes it to a projective method based on exact numerical quadrature, with a
  formal energy cutoff.}
\bibitem{note4} {In this and other simulations we have observed a tendency of
vortices to
  cluster together in (short-lived) co-rotating pairs which orbit their
  geometric centre, a behaviour characteristic of two-dimensional superfluid
  turbulence (see \cite{Wang07} and references therein).}
\bibitem{note5} {This energy change $\Delta E$ cannot be attributed to the
initial trap
  deformation associated with the stirrer, as the energy of an axially symmetry
  state is unchanged by the abrupt introduction of a perturbation of quadrupolar
  symmetry (i.e., within the sudden approximation \cite{Bransden00} implicit in
  our procedure).}
\bibitem{note7} {We note than in \cite{Holzmann07}, the fully bosonic analogue
of
  Eq.~(\ref{eq:fit_function1}) was derived as a self-consistent mean-field
  description of the total field density in the local density approximation as
  the BKT transition is approached from above.}
\bibitem{note8} {Condensates formed in dilute Bose gases may possess some
angular momentum
  without containing vortices due to their surface oscillations (see e.g.
  \cite{Stringari96}).}
\bibitem{note9} {It is important to note that such an artifact must arise in any
field theory
  which is truncated either explicitly or implicitly, and that the defect
  occurring here is thus close to the minimal defect attainable in such a
  description.}
\bibitem{note13} {While additional numerical procedures were used in
calculations presented in
  \cite{Parker06b,Parker05a}, these additions were not essential for lattice
  crystallization, i.e., relaxation to a lattice state was also observered in
  simulations performed without these additions \cite{note15}.}
\bibitem{note10} {We note that our calculations confirm that in the absence of
trap anisotropy
  and therefore without any dynamical instability, the presence of Wigner noise
  will eventually allow a condensate propagated in the rotating frame to
  nucleate vortices.}
\bibitem{note11} {An alternate (and contrary) explanation, that the field is
exhibiting
  quasi-recurrent behaviour in the manner of \cite{Fermi65} seems unlikely given
  that the trajectories appear to have converged to the temperatures predicted
  by the simple analysis presented in Sec.~\ref{subsec:Thermo_params}.}
\bibitem{note12} {At a late stage in the preparation of this manuscript we
became aware of
  experimental investigations into the nature of Bose-condensates in rotating,
  quasi-2D geometries \cite{Lamporesi08}.}
\end{thebibliography}

%%%%%%%%%%%%%%%%%%%%%%%%%%%%%%%%%%%%%%%%%%%%%%%%%%%%%%%%%%%%%%%%%%
\end{document}